\documentclass[aps,pre,twocolumn,groupedaddress]{revtex4}
\usepackage{amsmath}
\usepackage{graphicx}

\begin{document}
\title{Moving boundary approximation for curved streamer ionization fronts: Solvability analysis}
\author{Fabian Brau$^1$, Benny Davidovitch$^{2}$ and Ute Ebert$^{1,3}$ }
\affiliation{$^1$ Centrum Wiskunde \& Informatica (CWI), P.O.Box 94079, 1090 GB Amsterdam, The Netherlands.}
\affiliation{$^2$ Physics Department, University of Massachusetts, Amherst MA 01003} \affiliation{$^3$Dept. Physics, Eindhoven Univ.\ Techn., The Netherlands.}

\date{\today}

\begin{abstract}
The minimal density model for negative streamer ionization fronts is investigated. An earlier moving boundary approximation for this model consisted of a ``kinetic undercooling" type boundary condition in a Laplacian growth problem of Hele-Shaw type. Here we derive a curvature correction to the moving boundary approximation that resembles surface tension. The calculation is based on solvability analysis with unconventional features, namely, there are three relevant zero modes of the adjoint operator, one of them diverging; furthermore, the inner/outer matching ahead of the front has to be performed on a line rather than on an extended region; and the whole calculation can be performed analytically. The analysis reveals a relation between the fields ahead and behind a slowly evolving curved front, the curvature and the generated conductivity. This relation forces us to give up the ideal conductivity approximation, and we suggest to replace it by a constant conductivity approximation. This implies that the electric potential in the streamer interior is no longer constant but solves a Laplace equation; this leads to a Muskat-type problem.
\end{abstract}

\maketitle

\section{Introduction}
\label{sec1}

Streamers are growing plasma channels that in sufficiently strong electrostatic field expand into ionizable matter like air or other gases, liquids or solids. The field accelerates the electrons sufficiently to ionize gas molecules upon collisions. The self-focussing of the field at the tip of the propagating streamer strongly supports this process. The richness of spatio-temporal structures that form as streamers undergo successive branching events during their propagation make them an important example of nonequilibrium pattern formation. They are of fundamental interest not only for this reason, but also because they determine the initial stages of electric breakdown in sparks and lightning, in technological applications as well in natural processes~\cite{Raether,Loeb,Raizer,PSST}. In this paper we focus on negative (anode-directed) streamers in simple gases like pure nitrogen or argon. Their dynamics can be described by a minimal model of nonlinear reaction-advection-diffusion equations, as described, e.g., in \cite{PRLUWC,PREUWC}. The model describes the evolution of electron and ion densities and their nonlinear coupling to the local electric field.

\begin{figure}[!hbtp]
\centerline{\includegraphics[width=6cm,clip]{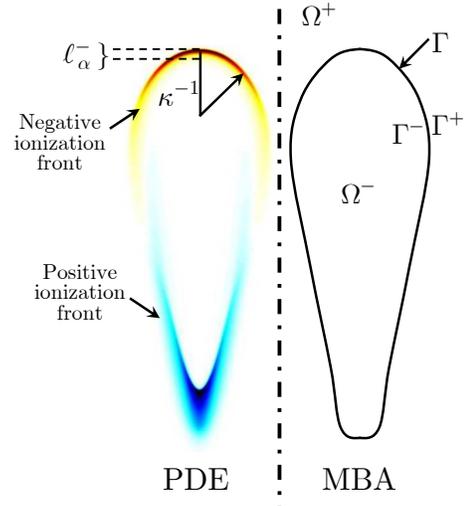}}
\caption{On the left: generic solution (net charge density) of the minimal PDE model with curvature $\kappa$ and width $\ell_{\alpha}^-$ (see Eq.~(\ref{lambdaminus1})) of the ionization front. On the right: the moving boundary approximation (MBA) with the ionized region, $\Omega^-$, the non-ionized region, $\Omega^+$, and the sharp interface, $\Gamma$.} \label{fig00}
\end{figure}

Many simulations~\cite{Kunhardt,dhal85,dhal87,vite94,PRLMan,AndreaRapid,CaroPRE,CaroJCP} since the
early 1980'ies have shown that the ionized interior of the streamer finger is preceded and surrounded by a narrow curved space charge layer that enhances the electric field in the non-ionized region ahead of the front and screens it in the ionized interior. In particular, in strong background fields after a sufficiently long evolution, the width of the ionization front can be much smaller than its radius of curvature~\cite{CaroPRE,CaroJCP,luqu07}. This separation of scales enables one to consider the front as an infinitesimally thin, sharp moving interface $\Gamma$. Many inhomogeneous systems involve domains of well defined phases separated by thin interfaces. These include non-equilibrium systems like solidification occurring by dendritic growth \cite{lang80,ben84,coll85,caro92,saleh97,karm98,ande01}, bacterial growth \cite{ben00,mull02} or many diffusive-reactive systems \cite{fife88}. In Fig.~\ref{fig00}, we present a generic solution for the net charge density of the minimal PDE model showing the separation of scale together with the resulting moving boundary approximation. The original nonlinear dynamics is then replaced by a set of linear field equations (typically Laplace) on both sides of $\Gamma$, $\Omega^+$ and $\Omega^-$, with appropriate boundary conditions at the interface, $\Gamma^+$ and $\Gamma^-$ respectively, and further away from it \cite{karm98,almg99,mcfa00,elde01} (and references therein); the nonlinearity enters through the motion of the boundary. The interface dynamics is then typically related to gradients of the Laplacian fields at its vicinity, $\Gamma^+$ and $\Gamma^-$. 

In the context of streamer dynamics, the concept of an interfacial approximation was proposed already in the early 1970'ies by Lozansky and Firsov before any numerical simulations~\cite{loza73}. At the time, their model was in competition with other streamer models and concepts. Lozansky and Firsov proposed to consider the streamer interior, $\Omega^-$, as ideally conducting, i.e., the electric potential $\phi$ to be constant in the interior. The exterior, $\Omega^+$,  is non-ionized and therefore does not contain space charges, solving in electrostatic approximation (that is justified in~\cite{PREUWC})
\begin{equation}
\label{1} \nabla^2\phi=0 \quad \text{in} \quad \Omega^+.
\end{equation}
The interface was assumed to move with the local electron drift velocity
\begin{equation}
\label{2}
\vec{v}= \nabla\phi^+.
\end{equation}
From here, superscripts $^\pm$ attached to fields, potential and densities indicate their limit value as they approach the interface from $\Omega^+$ and $\Omega^-$, respectively. In particular, we denote $\phi^+\equiv\phi|_{\Gamma^+}$ and $\phi^-\equiv\phi|_{\Gamma^-}$. However, the solutions of this interfacial model were hardly explored. Only 30 years later, the interfacial model was taken up again in the streamer context~\cite{meul04}, and Meulenbroek {\it et al.} showed that the
model actually allowed for spontaneous streamer branching. However, the model was also found to be
mathematically ill posed; in the context of fluid dynamics, this is explain for example in Ref.~\cite{howi86} and references therein. To resolve this problem, the boundary condition $\phi^+-\phi^-=0$ was replaced by the regularizing boundary condition
\begin{equation}
\label{GLkatze} \phi^+-\phi^-=Q_0(\hat{{\bf n}}\cdot \nabla\phi^+)\stackrel{|\hat{{\bf n}}\cdot \nabla\phi^+|\gg1}\longrightarrow \hat{{\bf n}}\cdot \nabla\phi^+
\end{equation}
which was proposed in~\cite{PRL05} and derived in planar front approximation in~\cite{brau07}. The boundary condition accounts for the finite width of the charged layer (the ionization front) that leads to a finite variation of the electric potential across the front. The boundary condition in the limit of large electric fields actually turns out to be identical to the ``kinetic undercooling" boundary condition that was applied to crystal growth under certain conditions \cite{kuik85,chap03}.

The solutions of the interfacial model with this boundary condition (\ref{GLkatze}) implemented were studied in~\cite{PRL05,SIAM06,brau07,saleh08}, and the results of these papers show that the interfacial model employed there is the simplest one that regularizes the motion -- therefore we call this model the minimal regularized model. However, it also was shown recently that this analysis applies only to boundaries that at time $t=0$ are sufficiently many times differentiable~\cite{Prok08}.

In this paper we compute curvature corrections to the boundary condition (\ref{GLkatze}). A systematic expansion of slightly curved fronts about planar fronts for streamers was first suggested in~\cite{PRLUWC}, and \cite{PREUWC} contains the analysis of planar fronts as first step in this research program. In analogy with other weakly curved fronts, a perturbative expansion of the curved front about the planar front with subsequent solvability analysis was the method at hand. Solvability analysis means that the inner front structure is integrated out and replaced by conditions for interfacial motion that are matched to the dynamics on the outer scale.

However, such an analysis could not be performed on the streamer model of~\cite{PRLUWC,PREUWC} as the fronts are pulled; for pulled fronts, we refer to~\cite{Pulled98,Pulled00}, and for the impossibility of a solvability analysis for pulled fronts to~\cite{Sol}. The formal reason for the non-applicability of the method are diverging integrals in the leading edge of the ionization front; the physical reason is the algebraically slow front relaxation of the leading edge that actually stretches out through the whole non-ionized region and does not make part of the inner front region that is to be integrated out. The pulled nature of the front also requires special care with grid refinement in numerical solutions~\cite{CaroJCP}.

However, the leading edge that pulls the front, is diffusive, and it is a physically and mathematically meaningful approximation to neglect electron diffusion in strong fields, where electron motion due to drift dominates over the diffusive motion~\cite{Man04,gianne}. We have employed this approximation in the interfacial model and we have checked that its results approximate simulations with electron diffusion well~\cite{brau07,ST08}. We here employ the same approximation which now makes it possible to perform a solvability analysis.

In fact, neglecting electron diffusion and assuming that the state ahead of the front is completely
non-ionized, converts the inner-outer matching problem from one non-typical situation to the other. With electron diffusion, the inner front region was not separated from the outer one. Without electron diffusion, however, the inner region finishes precisely where the electron density discontinuously jumps to 0, and the matching between inner front region and the non-ionized outer region takes place precisely on this line rather than on an extended spatial region.

The ``moving boundary condition'' (MBC) for a slightly curved front dynamics now can be systematically derived from the original nonlinear field equations: A perturbation of a planar front is assumed whose curvature in the direction transverse to the front motion is much smaller than the front width, and solvability analysis is used to connect the perturbed values of the fields ahead and behind the curved front. Recently, such approach has been successfully applied to derive MBC equations for the dynamics of discontinuity curves which appear in nonlinear diffusion problems \cite{mull02}.

The solvability analysis poses technical challenges. We will show that the electrostatic field behind a slowly evolving curved front does not vanish but is rather constant and proportional to the front curvature. Consequently, the electric potential diverges linearly behind the front. Another divergence appears in one of the zero modes of the solvability analysis. This divergence is however necessary to cancel the one of the electric potential behind the front in order to get a meaningful boundary condition for the jump in the electric potential across the interface~$\Gamma$. To our knowledge, this nontrivial computational aspect is new, and we are not aware of similar types of solvability analysis in implementations of sharp interface approach to front dynamics in other physical systems.

Furthermore, since the potential is no longer constant behind the front, the ideal conductivity approximation ($\phi=0$ in $\Omega^-$) must be relaxed and is replaced by a constant conductivity approximation for the ionized streamer interior ($\nabla^2 \phi=0$ in $\Omega^-$). 

Explicitly, our front dynamics consists of the following equations:
\begin{eqnarray}
\Delta \phi &=& 0 \quad \text{in} \quad \Omega^+ \label{laplaceaheadintro} \\
\Delta \phi &=& 0 \quad \text{in} \quad \Omega^- \label{laplaceouterintro}
\end{eqnarray}
with the MBC
\begin{eqnarray}
\hat{{\bf n}} \cdot \nabla \phi^- &=& Q_2(\hat{{\bf n}} \cdot \nabla \phi^+)
\kappa \label{BCphi-intro} \\
\phi^+-\phi^- &=& Q_0(\hat{{\bf n}} \cdot \nabla \phi^+) + Q_1(\hat{{\bf n}}
\cdot \nabla \phi^+) \kappa \label{BCpotjumpintro} \\
v_{n} &=& \hat{{\bf n}} \cdot \nabla\phi^+ \label{velocityfrontintro}.
\end{eqnarray}where the coefficients $Q_i$ depend on the electrostatic field ahead the front,
and are given by analytic formulas derived from the planar front solution, see Eqs.~(\ref{q0})-(\ref{q2}).

The paper is organized as follows: In Sec.~\ref{collection} we summarize previous results on the minimal model and its planar front solutions essential for our analysis. In Sec.~\ref{pertubation}, we start our formal derivation by assuming a slightly curved front with $\epsilon =\kappa \ell_{\alpha}^{-} \ll 1$, where $\ell_{\alpha}^{-}$ and $\kappa$ are the width and the curvature of the front respectively. We construct a perturbative expansion of the solution to the minimal model dynamics in powers of $\epsilon$ around the zeroth order planar front solution. We focus on the leading order behavior in $\epsilon$ and describe it as a solution to a set of four inhomogeneous linear ODEs. In Sec.~\ref{sol-analysis}, we explain and develop the solvability analysis formalism: Using zero modes of the adjoint linear differential operator we derive relations between deviations of the electrostatic field and ion/electron densities ahead and behind the curved front, $\Gamma^+$ and $\Gamma^-$, from their values at the planar front solution. These relations turn out to be the required MBC equations for the curved fronts dynamics. The required zero modes are calculated analytically in Sec.~\ref{sec:exact}. In Sec.~\ref{sec:mbc} we conclude our inner analysis and derive the MBC equations. The final form of the MBC equations (\ref{laplaceahead})-(\ref{velocityfront}), obtained by matching inner and outer regions, is derived in Sec.~\ref{sec:refine} where we give also the large field limit of the interfacial model together with an elementary algorithm to solve it. In Sec.~\ref{CCApprox}, we discuss the constant conductivity approximation for the ionized region. In Sec.~\ref{conclusion} we conclude by highlighting the major results of this paper and point to future directions.

\section{Collection of some previous results}
\label{collection}

In this section we summarize previous results required for the present analysis. We briefly describe the minimal model, we discuss the front velocity and the coupling between the front and the non-ionized exterior region, and we cite several properties of planar uniformly translating fronts required for our further analysis.

\subsection{The minimal streamer model}
\label{minimal}

The model for negative streamers in simple non-attaching gases like nitrogen and argon as used, e.g.,
in~\cite{dhal87, vite94, PRLUWC, PRLMan, AndreaRapid, CaroPRE, CaroJCP, brau07} consists of a set of three coupled partial differential equations for the electron density $\sigma$, the ion density $\rho$ and the electric field ${\bf E}$. In dimensionless units the model reads
\begin{eqnarray}
\label{PDE1}
\partial_t \sigma - \nabla \cdot (\sigma{\bf E})&=&\sigma|{\bf E}| \;\alpha(|{\bf E}|),
\\
\label{PDE2}
\partial_t \rho &=& \sigma|{\bf E}| \;\alpha(|{\bf E}|),
\\
\label{PDE3}
\nabla \cdot {\bf E}&=&\rho-\sigma, \quad {\bf E} = -\nabla \phi.
\end{eqnarray}
A general discussion about dimensions for this model can be found, e.g., in \cite{PSST,PRLUWC,PREUWC,luqu07}. The first two equations are the continuity equations for the electrons and the ions while the last is the Coulomb equation for the electric field generated by the space charge $\rho-\sigma$ of electrons and ions. Here the electrostatic approximation for the electric potential $\phi$ was introduced~\cite{PREUWC}. Notice that the electron particle current is here taken as the drift current only $-\sigma{\bf E}$, neglecting electron diffusion. For the effect of a diffusive contribution, see the discussion in the introduction as well as~\cite{gianne}. The ion particle current is neglected because due to their larger mass the ions are much less mobile.

The term $\sigma |{\bf E}| \alpha(|{\bf E}|)$ is the generation rate of additional electron ion-pairs; it is the product of the absolute value of the drift current times the effective cross section $\alpha$ taken as field dependent; a commonly used form is the Townsend approximation
\begin{equation}
\label{alphaTown} 
\alpha(|{\bf E}|)=e^{-1/|{\bf E}|}.
\end{equation}
However, our analysis holds for the more general case of
\begin{eqnarray}
\label{alphagen}
\alpha(0)&=&0, \quad \frac{d\alpha(x)}{d{x}} \ge0, \nonumber \\
\alpha(x)&\underset{x\gg1}{=}&1-{\cal O}\left(\frac1 {x}\right).
\end{eqnarray}

\subsection{Ionization fronts and their velocity}
\label{planar}

We consider planar ionization fronts propagating in the positive $x$-direction into a medium that is
completely non-ionized beyond a certain point $x_f(t)$
\begin{eqnarray}
&\sigma=0=\rho & \quad \text{for} \quad x>x_f(t),\nonumber \\
&\sigma>0& \quad \text{for} \quad x<x_f(t),
\end{eqnarray}
and we work in the comoving coordinate
\begin{equation}
\label{xi} \xi=x-x_f(t).
\end{equation}
As is easily verified by rewriting Eq.~(\ref{PDE1}) in the comoving coordinate $\xi$ with the help of
(\ref{PDE3}) as
\begin{equation}
\partial_t\Big|_\xi \sigma - \left(\frac{\partial x_f}{\partial t}\;\hat{\bf x}+{\bf E}\right)\cdot\nabla\sigma =\sigma\;\left(\rho-\sigma+|{\bf E}| \;\alpha(|{\bf E}|)\right),
\end{equation}
where $\hat{\bf x}$ is the unit vector in the $x$-direction, the velocity of the front is determined purely by the value of the field $E^+$ on the discontinuity of the electron density
\begin{equation}
\label{vf} v_f(t)=\frac{\partial x_f(t)}{\partial t}=-E\left(x_f(t),t\right)\equiv E^+.
\end{equation}
Notice that, in order to simplify notations later, $E^+$ is a positive quantity.
Obviously, this relation between front velocity and local field is not restricted to planar fronts only, but holds generally:
\begin{equation}
\label{fieldv} {\bf v}_f(t)=\frac{\partial {\bf r}_f(t)}{\partial t}=-{\bf E}\left({\bf r}_f(t),t\right).
\end{equation}
As the space charge $\rho-\sigma$ never diverges, the field ${\bf E}$ is continuous across ${\bf r}_f$.

\subsection{Coupling the front to the outer non-ionized region}
\label{phi+}

In the purely one-dimensional planar setting, the field does not vary ahead of the front: $\partial_x E=0$ for $x>x_f(t)$. On the other hand, the curved ionization front around the tip of a streamer finger creates the characteristic field enhancement ahead of the streamer tip. However, the region $x>x_f(t)$ can be left purely to the outer scale analysis of the Laplace equation
\begin{equation}
\label{Laplace+} \nabla^2\phi=0,\quad {\bf E}=-\nabla\phi \quad \text{for} \quad \xi>0
\end{equation}
and has no further influence on the front solution except for determining the field ${\bf E}^+$ at ${\bf r}_f(t)$. The matching between inner and outer region here is completely concentrated on the line ${\bf r}_f(t)$. As also remarked already in the introduction, these two spatial regions only decouple if the electron diffusion $D$ (that is included in numerical streamer models) is neglected. Putting $D=0$ eliminates the leading edge and the pulled dynamics of the front \cite{PREUWC,Man04}.

\subsection{Uniformly translating planar ionization fronts}
\label{UTF}

If $E^+$ is time independent, $\partial_t E^+=0$, the front propagates uniformly with velocity $v=E^+$.
Uniformly translating front solutions of the minimal model~(\ref{PDE1})--(\ref{PDE3}) in a field $E^+$
depend only on the comoving coordinate $\xi = x - v t$. They have been discussed in many
previous papers~\cite{PRLUWC,PREUWC,Man04}, for a recent thorough discussion, we refer, in particular,
to section II of~\cite{brau07} and for the comparison of fronts with or without electron diffusion $D$ to section 2 of~\cite{gianne}.

\begin{figure}[!hbtp]
\includegraphics[width=\columnwidth,clip]{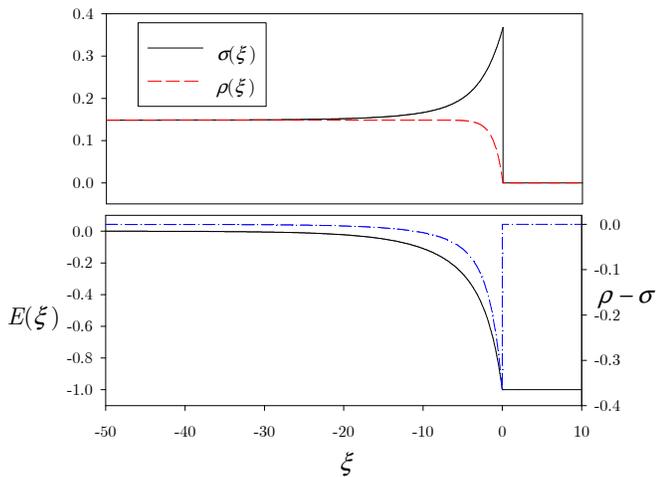}
\caption{Densities and field for a uniformly translating planar ionization front in a far field
$|E_{\infty}|=E^+=1$ as a function of the comoving coordinate $\xi$ (\ref{xi}).
The front moves to the right with velocity $v=E^+$. Upper panel: electron
density $\sigma$ (solid line) and ion density $\rho$ (dashed line); lower panel:
electric field (solid line, axis on the left) and space charge density $\rho-
\sigma$ (dashed-dot line, axis on the right). For $\alpha(|{\bf E}|)$ the Townsend
approximation (\ref{alphaTown}) was used.}
\label{fig02}
\end{figure}

Analytically the front solution in the ionized region $\xi<0$ is given implicitly by the equations
\begin{eqnarray}
\label{sigma}
\sigma(E)&=&\frac{E^+}{E^+ -|E|}\;\rho(E),\\
\label{rho}
\rho(E)&=&\int_{|E|}^{E^+}d\mu \, \alpha(\mu) ,\\
\label{E} 
-\xi&=&\int_{E(\xi)}^{-E^+}\frac{d\mu}{\mu} \frac{E^+ +\mu}{\rho(\mu)}.
\end{eqnarray}
Fig.~\ref{fig02} shows the shape of the solution for $E^+=1$. At the shock front, the electron density jumps from zero to
\begin{equation}
\label{edensjump} \sigma^+ \equiv\lim_{\xi\uparrow0}\sigma(\xi)=E^+\, \alpha(E^+),
\end{equation}
while ion density and electric field are continuous across the interface
\begin{equation}
\rho(\xi)={\cal O}(\xi),\quad E(\xi)=-E^+ +{\cal O}(\xi)\quad \text{for} \quad \xi<0.
\end{equation}
For $\xi\rightarrow - \infty$ the densities approach the value
\begin{eqnarray}
\label{edensminfty} 
\sigma^-&\equiv& \sigma(-\infty)=\int_0^{E^+}d\mu\, \alpha(\mu),\\
\label{rholimit} 
\rho^-&\equiv& \rho(-\infty)=\sigma^-,
\end{eqnarray}
which means that the space charge density $\rho-\sigma$ vanishes, and the electric field vanishes as well
\begin{equation}
\label{Elimit} 
E^-\equiv E(-\infty)=0.
\end{equation}

Far behind the front, where ${\bf E}$ is so small that $\alpha(|{\bf E}|)\simeq 0$, the profiles of {\bf E}
and $\sigma$ decay exponentially:
\begin{equation}
    \label{asympplanar}
    \left(\begin{array}{c}\sigma\\ \rho\\ E
\end{array}\right)(\zeta)=\left(\begin{array}{c}\sigma^-\\ \sigma^-\\ 0
\end{array}\right)+A\left(\begin{array}{c}\lambda\\ 0\\ 1
\end{array}\right)e^{\lambda\, \zeta}+O\left(e^{2\lambda \, \zeta}\right),
\end{equation}
where
\begin{equation}
\label{lambdaminus}
\lambda=\frac{\sigma^-}{v}=\frac{1}{E^+}\int_0^{E^+}d\mu\, \alpha(\mu)
\stackrel{E^+\to \infty}{\longrightarrow} 1.
\end{equation}
For more details on the asymptotic behavior at $\xi=0$ and $\xi\to-\infty$, we refer to section II
of~\cite{Man04}.

The exponential decay in (\ref{asympplanar}) suggests a natural definition of the front width as
\begin{equation}
    \label{lambdaminus1}
    \ell_\alpha^-=\frac1\lambda=\frac{v}{\sigma^-}.
\end{equation}
The length scales within the front are further elaborated in section 6.3.2 of~\cite{gianne} where next to $\ell_\alpha^-$ also another scale $\ell_\alpha^+=1/\alpha(E^+)$ was defined that characterizes the inner front region close to $\xi=0$ for $D=0$. (Nonvanishing electron diffusion $D$ introduces another scale $\ell_D$.) The length scale $\ell_\alpha^+$ is obviously of no relevance for matching of the outer scale, but $\ell_\alpha^-$ is the appropriate quantity.

From the uniformly translating planar front solution, the boundary condition (\ref{GLkatze}) was derived in~\cite{brau07} as follows. Integration of the electric potential, $\partial_{\zeta}\phi(\zeta)=-E(\zeta)$,
from $-\infty$ to 0 leads to
\begin{equation}
    \phi(0)-\phi(-\infty)\equiv \phi^+ -\phi^-=-\int_{-\infty}^0 d\zeta \,
E(\zeta).
\end{equation}
Substituting the variable $E(\zeta)=-x$ with the help of~(\ref{E}), and using the limit (\ref{Elimit}) for
the lower limit of integration, we obtain
\begin{equation}
    \label{defq0}
    \phi^+ -\phi^-=Q_0(E^+),\quad Q_0(E^+)\equiv \int_0^{E^+} dx\, \frac{E^+ -x}{\rho(x)},
\end{equation}
where the function $\rho(x)$ is defined in~(\ref{rho}). As discussed in detail in~\cite{brau07}, the
function $Q_0$ depends nonlinearly on $E^+$ (see Fig.~\ref{fig05}), but asymptotically
\begin{equation}
\label{phiLim} \phi^+ -\phi^-\to E^+ \quad \text{for} \quad E^+ \gg 1.
\end{equation}

Finally, let us also recall an important property of the planar front solution which is a useful relation used later in the paper
\begin{equation}
v \frac{dE(\xi)}{d\xi} - \sigma(\xi) E(\xi) = 0. \label{chargeplanar}
\end{equation}
This relation is obtained by taking the derivative of Eq.~(\ref{E}) and using Eq.~(\ref{sigma}).

\subsection{The minimal regularized boundary model}
\label{MRBM}

The moving boundary model derived and evaluated in~\cite{PRL05,SIAM06,brau07,saleh08} is given by
Eqs.~(\ref{1})--(\ref{GLkatze}) together with the ideal conductivity approximation $\phi={\rm const.}$ in $\Omega^-$ that will further be discussed in Sec.~\ref{CCApprox}. This model can be considered as a minimal regularized boundary model. While the model with the boundary condition $\phi^+-\phi^-=0$ treated in \cite{meul04} lacks regularization and for generic initial conditions breaks down within infinitesimal time, the model of~\cite{PRL05,SIAM06,brau07,saleh08} according to our analysis does stay regularized as long as the initial contour is infinitely many times differentiable. It consists of the exact description of the non-ionized region~(\ref{Laplace+}), of the exact relation between velocity and local field~(\ref{fieldv}) and furthermore of two approximations for the full dynamics of the PDE's~(\ref{PDE1})--(\ref{PDE3}). First, the jump of the electric potential across the front is modeled using the approximation~(\ref{phiLim}) of a planar uniformly translating front in the limit of large field ${\bf E}^+$; corrections to this behavior will be studied through the solvability analysis from section~\ref{sec:BCweakly} on. Second, the streamer interior is modeled as ideally conducting: $\phi={\rm const.}$ in $\Omega^-$. This approximation will be discussed and improved in Sec.~\ref{CCApprox}.

\section{Boundary approximation for slightly curved fronts}
\label{sec:BCweakly}

To derive curvature corrections to the minimal regularized boundary model (section~\ref{MRBM}) from the underlying PDE's described in section~\ref{minimal}, we follow the general approach as described, e.g., in \cite{karm98,mull02,almg99,mcfa00,elde01,Sol,folc99} for solidification fronts in supercooled melts, nonlinear diffusion fronts or interface between two inmiscible fluids, and adapt it to our problem as necessary.

\subsection{Perturbative expansion about planar fronts}
\label{pertubation}

In a first step, we expand slightly curved fronts about the uniformly translating planar fronts that were recalled in section~\ref{UTF}. Weak curvature means that the width of the front $\ell_\alpha^-$
(\ref{lambdaminus1}) is much smaller than the mean radius of curvature $R=1/\kappa$ of the front. The quantity
\begin{equation}
	\epsilon=\kappa \ell_\alpha^-
\end{equation}
is then a small parameter. The inner region of the curved front can be expanded about the planar front by introducing a curvilinear orthogonal coordinate system, with coordinates $(\eta,\zeta)$ locally tangential and normal to the moving front. We expand the inner region as
\begin{eqnarray}
\phi(\eta,\zeta,t) &=& \phi_0(\zeta) + \epsilon \phi_1 + \cdots \nonumber \\
\sigma(\eta,\zeta,t) &=& \sigma_0(\zeta) + \epsilon \sigma_1 + \cdots \nonumber \\
\rho(\eta,\zeta,t) &=& \rho_0(\zeta) + \epsilon \rho_1 + \cdots \nonumber \\
v(\eta,t) &=& v_0 + \epsilon v_1 + \cdots , \label{expansion}
\end{eqnarray}
where $[\phi_0(\zeta), \sigma_0(\zeta),\rho_0(\zeta)]$ and $v_0=E^+$ are the planar uniformly translating front solutions (\ref{sigma})-(\ref{E}) within a fixed field $E^+$.

Assuming that the variations in the transversal direction $\eta$ are of the order of the radius of curvature $R$, it is useful to rescale the transversal coordinate as $\tilde\eta = \epsilon\eta$. A divergence $\nabla \cdot {\bf  J}$ then can be expanded in $\epsilon$ as
\begin{equation}
\nabla  \cdot {\bf J}  = \frac{\partial}{\partial \zeta} J_\zeta +
\epsilon \frac{\partial}{\partial \tilde{\eta}} J_{\tilde{\eta}},
\label{divergence}
\end{equation}
and the Laplacian $\nabla^2 \phi$ as well:
\begin{eqnarray}
\nabla^2  &=& \frac{\partial^2}{\partial\zeta^2} + \kappa
\frac{\partial}{\partial\zeta} + \epsilon^2
\frac{\partial^2}{\partial\tilde{\eta}^2}, \nonumber \\
\label{laplacian} &=& \frac{\partial^2}{\partial\zeta^2} + \epsilon \lambda \frac{\partial}{\partial\zeta} +
O(\epsilon^2), \quad \lambda=\frac1{\ell_\alpha^-}.
\end{eqnarray}
Therefore, the equations for $[\phi_1,\sigma_1,\rho_1]$ and $v_1$ depend only on $\zeta$ as common. 

Substituting the expansion (\ref{expansion}) in the minimal model equations (\ref{PDE1})-(\ref{PDE3}) we find to ${\cal O}(\epsilon^0)$ the planar front Eqs.~(\ref{sigma})-(\ref{E}). We obtain to ${\cal O}(\epsilon)$ the equations:
\begin{eqnarray}
\label{perturbode}
(\rho_0-2\sigma_0+f_0)\;\sigma_1+\sigma_0\;\rho_1+(\partial_{\zeta}\sigma_0-
\sigma_0 f_0')E_1 \nonumber \\ +(v_0+E_0)\;\partial_{\zeta}\sigma_1 = -
v_1\partial_{\zeta}\sigma_0, \label{expansion1}\\
f_0\;\sigma_1-\sigma_0 f_0'E_1+ v_0\;\partial_{\zeta}\rho_1 = -
v_1\partial_{\zeta}\rho_0, \label{expansion2}\\
-\sigma_1+\rho_1-\partial_\zeta E_1 = \lambda E_0, \label{expansion3} \\
E_1 = -\partial_{\zeta} \phi_1, \label{expansion4}
\end{eqnarray}
where we used
\begin{equation}
f(|E_0+E_1|)=f_0 - E_1 f_0',
\end{equation}
and where
\begin{eqnarray}
    \label{deff0}
    f_0&=&f(|E_0|)=|E_0|\alpha_0=|E_0|\alpha(|E_0|), \\
    \label{deff0p}
    f_0'&=&d_x f(x)|_{x=|E_0|}=\alpha_0(1-1/E_0).
\end{eqnarray}
These equations are linear for $(\sigma_1,\rho_1, E_1,\phi_1)$ with inhomogeneous terms $v_1\partial_\zeta\sigma_0$, $v_1\partial_\zeta\rho_0$ and $E_0$. The structure of this system of ODEs reads
\begin{eqnarray}
\label{linearsystem}
&&{\cal L}\cdot{\bf U}_1= {\cal N}, \quad{\cal L}={\bf M}-{\bf D}
\partial_\zeta,
\\
\label{vectorU1andI}
&&{\bf U}_1=
\left(\begin{array}{c}\sigma_1\\ \rho_1\\ E_1\\ \phi_1 \end{array}\right), \quad
{\cal N}=\left(\begin{array}{c}-v_1\partial_{\zeta}\sigma_0\\
                        -v_1\partial_{\zeta}\rho_0\\ \lambda E_0 \\0
        \end{array}\right),
\\
\label{matrixM}
&&{\bf M}(\zeta)=\left(\begin{array}{cccc}
  \rho_0-2\sigma_0+f_0 & \sigma_0 & \partial_{\zeta}\sigma_0-\sigma_0 f_0' & 0
\\
  &&\\
  f_0                  &     0    & -\sigma f_0'                           & 0
\\
  &&\\
  -1                   &     1    &     0                                  & 0
\\
  &&\\
  0                    &     0    &     -1                                 & 0
\\
\end{array}\right),
\\
\label{matrixD}
&&{\bf D}(\zeta)=\left(\begin{array}{cccc}
  -(v_0+E_0) & 0     & 0  & 0\\
  &&\\
  0          & -v_0  & 0  & 0\\
  &&\\
  0          &  0    & 1  & 0\\
  &&\\
  0          &  0    & 0  & 1\\
\end{array}\right).
\end{eqnarray}

\subsection{The structure of the solvability analysis for streamer fronts}
\label{sol-analysis}

In this subsection, using the zero modes of the adjoint operator ${\cal L}^*$ of ${\cal L}$, we formally derive corrections to the boundary relations derived from the planar front solution. The explicit expressions of the coefficients in these relations are calculated in later subsections.

\subsubsection{Formal procedure of solvability analysis}

The boundary conditions are evaluated through a solvability analysis. We first introduce its general
formalism. Consider a system of inhomogeneous linear ODEs
\begin{equation}
\label{57}
    {\cal L}\cdot{\bf U}_1={\cal N},
\end{equation}
where ${\cal L}$ is a linear differential operator that can be written as
\begin{equation}
    \label{operator}
    {\cal L}={\bf M}-{\bf D}\partial_{\zeta},
\end{equation}
and where {\bf M} and {\bf D} are matrices whose entries depend on $\zeta$. The adjoint operator of ${\cal L}$ is given by
\begin{equation}
    \label{adjoperator}
    {\cal L}^*={\bf M}^t +\partial_\zeta({\bf D}^t) + {\bf D}^t \partial_\zeta,
\end{equation}
where $^t$ stands for the transpose of the matrix. Assume that a zero mode, ${\bf U}^*$, of ${\cal L}^*$ is known
\begin{equation}
    \label{zeromodeeq}
{\cal L}^* \cdot{\bf U}^*=0 \quad \text{on} \quad a\le\zeta\le b.
\end{equation}
Multiplying Eq.~(\ref{57}) with $({\bf U}^*)^t$, integrating over $[a,b]$ and integrating by part, one easily verifies that
\begin{equation}
\label{linear} \int_a^b d\zeta \;\left({\bf U}^*\right)^t {\cal N} = - \left[\left({\bf U}^*\right)^t {\bf
D} {\bf U}_1\right]_a^b.
\end{equation}
Here the arbitrary integration boundaries $a$ and $b$ were introduced to control possible divergences of the integral and the boundary term. The relation (\ref{linear}) is the basis of the solvability analysis performed in this work. As we will show in the next sections, taking the limits $a \to -\infty$ and $b \to 0$, relation (\ref{linear}) yields linear relations between the perturbed electrostatic field and potential on both ends of the inner region, thus providing boundary conditions for the Laplace equations (\ref{laplaceaheadintro}) and (\ref{laplaceouterintro}) in the outer regions.

It should be remarked that the different coupling to the outer regions on both sides of the front is
reflected by the asymmetric integration boundaries. At $\zeta=0$, there is an abrupt transition from the inner front region to the outer non-ionized region as described in section~\ref{phi+}; here the matching region shrinks to a line. For $\zeta\to -\infty$, there is an exponential decay to charge neutrality and the outer ionized region as discussed later in section~\ref{CCApprox}.

\subsubsection{Zero modes of ${\cal L}^*$ and MBC equations}
\label{zeromodesmbc}

Denoting the components of the zero modes as
\begin{equation}
    \label{ustar}
    {\bf U}^*=\left(\begin{array}{c}\psi_{\sigma}\\ \psi_{\rho}\\ \psi_E\\
\psi_{\phi}\end{array}\right),
\end{equation}
relation (\ref{linear}), with $a=\zeta_{\text{c}}$ and $b=0$, leads to the relation
\begin{eqnarray}
    \label{BC}
    &&\left[(v_0+E_0)\sigma_1\psi_{\sigma} +v_0\rho_1 \psi_{\rho}-\psi_E E_1-
\psi_{\phi} \phi_1\right]_{\zeta_{\text{c}}}^0 = \nonumber \\ &&-A v_1+ B
\lambda,
\end{eqnarray}
where
\begin{eqnarray}
    \label{eqA}
    A&=&\int_{\zeta_{\text{c}}}^0 d\zeta \left[ \psi_{\sigma}
\partial_{\zeta}\sigma_0 +\psi_{\rho} \partial_{\zeta}\rho_0 \right], \\
    \label{eqB}
    B&=&\int_{\zeta_{\text{c}}}^0 d\zeta\, E_0\, \psi_E.
\end{eqnarray}
In order to obtain the MBC equations we have to consider the limit $\zeta_{\text{c}} \to -\infty$. This limit has to be analyzed with caution due to the mutual cancellation of diverging terms. In order to extract from Eq.~(\ref{BC}) appropriate MBC equations, we use some properties of the perturbed field and densities. First, from Eq.~(\ref{fieldv}) and the expansion (\ref{expansion}) we have
\begin{equation}
    \label{v1}
    v_1=-E_1(0).
\end{equation}
Second, asymptotic analysis of Eq.~(\ref{perturbode}) for $\zeta \to -\infty$
leads to
\begin{equation}
    \label{sig1=rho1}
    \sigma_1(-\infty)=\rho_1(-\infty),
\end{equation}
which ensures charge neutrality at $\zeta = -\infty$ to order $\epsilon$. Third, we show below that $\psi_{\sigma}(0)$ does not diverge. Since from Eq.~(\ref{fieldv}), $v_0+E_0$ vanishes at $\zeta=0$, the finite value of $\sigma_1(0)$ does not enter the MBC equations. Finally, since $\rho$ is a
continuous function and $\rho_0$ vanishes for $\zeta>0$ (\ref{rholimit}), $\rho_1(0)$ must vanish as well:
\begin{equation}
    \rho_1(0)=0.
\end{equation}
With these relations, Eq.~(\ref{BC}) takes a simpler form
\begin{eqnarray}
    \label{BC2}
    &-&[A+\psi_E(0)]v_1+B \lambda
=E_1(\zeta_{\text{c}})\psi_E(\zeta_{\text{c}})
+\psi_{\phi}(\zeta_{\text{c}})\phi_1(\zeta_{\text{c}}) \nonumber \\ &-
&\psi_{\phi}(0)\phi_1(0)-
v_0\sigma_1(\zeta_{\text{c}})[\psi_{\sigma}(\zeta_{\text{c}})+\psi_{\rho}(\zeta_
{\text{c}})],
\end{eqnarray}
where we assumed $|\zeta_{\text{c}}| \gg \ell_\alpha^-$, such that $E_0(\zeta_{\text{c}})$,
$\sigma_0(\zeta_{\text{c}})$ and $\rho_0(\zeta_{\text{c}})$ reach their plateau values (\ref{asympplanar}).

In order to extract the actual MBC from Eq.~(\ref{BC2}), we must find the zero modes of ${\cal L}^*$. The number of independent MBC equations is related to the number of zero modes that give rise to converging integrals in the expressions for the coefficients $A,B$ (see Eqs.~(\ref{eqA}) and (\ref{eqB})). The calculation of these zero modes and the consequent evaluation of the required integrals are the subject of this section and the next one.

\subsubsection{The asymptotic behavior of zero modes at $\zeta \to -\infty$}
The construction of the adjoint operator ${\cal L}^*$ from Eq.~(\ref{adjoperator}) is straightforward. The system of equations for the zero modes, Eq.~(\ref{zeromodeeq}), is:
\begin{eqnarray}
\label{eq0mode1b}
(f_0-\sigma_0)\psi_{\sigma}+f_0\psi_{\rho} -\psi_E -
(v_0+E_0)\partial_{\zeta}\psi_{\sigma}&=&0, \\
\label{eq0mode2b}
\sigma_0\psi_{\sigma}+\psi_E-v_0\partial_{\zeta}\psi_{\rho}&=&0, \\
\label{eq0mode3b}
(\partial_{\zeta}\sigma_0-\sigma_0 f_0')\psi_{\sigma}-\sigma_0 f_0' \psi_{\rho}
- \psi_{\phi}+ \partial_{\zeta}\psi_E&=&0, \\
\label{eq0mode4b}
\partial_{\zeta} \psi_{\phi}&=&0.
\end{eqnarray}
Equation (\ref{eq0mode4b}) immediately gives
\begin{equation}
\label{psiphi}
\psi_{\phi}=c_{\phi}=\text{const}.
\end{equation}
As we explained above, the zero modes which are relevant for the derivation of MBC equations are those that give rise to converging integrals in Eqs. (\ref{eqA}) and (\ref{eqB}). It is thus essential to analyze first their asymptotic behavior in the limit $\zeta \to -\infty$. In this limit,
$\sigma_0=\rho_0=\sigma^-$ and $E_0=f_0=f'_0=0$, and the system of Eqs.~(\ref{eq0mode1b})-(\ref{eq0mode4b}) reduces to
\begin{eqnarray}
\label{asymeq1}
-\sigma^-\psi_{\sigma}-\psi_E-v_0\partial_{\zeta}\psi_{\sigma}&=&0, \\
\label{asymeq2}
\sigma^-\psi_{\sigma}+\psi_E-v_0\partial_{\zeta}\psi_{\rho}&=&0, \\
\label{asymeq3}
\partial_{\zeta}\psi_E&=&c_{\phi}.
\end{eqnarray}
Equation (\ref{asymeq3}) immediately gives:
\begin{equation}
\label{asympsie} \psi_E=c_{\phi}\zeta+ c_E \ ,
\end{equation}
where $c_E$ is another constant. Substituting Eq. (\ref{asympsie}) in (\ref{asymeq1}) and integrating we obtain
\begin{equation}
\label{asymppsis}
\psi_{\sigma}=c_{\sigma}e^{-\frac{\sigma^-}{v_0}\zeta}-
\frac{c_E}{\sigma^-}+\frac{c_{\phi}v_0}{(\sigma^-)2}-\frac{c_{\phi}}{\sigma^-
}\zeta \ ,
\end{equation}
where $c_{\sigma}$ is a third constant of integration. Finally, the asymptotic expression for $\psi_{\rho}$ is obtained by substituting Eqs. (\ref{asympsie}) and (\ref{asymppsis}) into (\ref{asymeq2}) and integrating:
\begin{equation}
\label{asymppsir}
\psi_{\rho}=-c_{\sigma}e^{-\frac{\sigma^-}{v_0}\zeta}+\frac{c_{\phi}}{\sigma^-}\zeta+c_{\rho} ,
\end{equation}
where the fourth constant of integration $c_{\rho}$ was introduced. The set of four independent constants $c_{\sigma}, c_{\rho}, c_E, c_{\phi}$ corresponds to four independent solutions of the set of four ODE's (\ref{eq0mode1b})-(\ref{eq0mode4b}).

The exponential divergence of the coefficient of $c_\sigma$ in equation (\ref{asymppsis}) and the
exponentially decaying profile of $\sigma_0(\zeta)$ with the same length scale (\ref{asympplanar}) imply that the integral (\ref{eqA}) converges only if $c_{\sigma} = 0$. This condition reduces the number of relevant zero modes, and hence the number of MBC equations derived from Eq. (\ref{BC2}), to three.

\subsubsection{Initial conditions for zero modes at $\zeta \to-\infty$}

In order to extract the three relevant zero modes from Eqs.~(\ref{eq0mode1b})-(\ref{eq0mode4b}) in the whole interval $\zeta \in (-\infty,0)$, it is natural to specify initial conditions $[\psi_{\sigma}(\zeta_{\text{ini}}), \psi_{\rho}(\zeta_{\text{ini}}), \psi_{E}(\zeta_{\text{ini}}),
 \psi_{\phi}(\zeta_{\text{ini}})]$ at some $\zeta=\zeta_{\text{ini}} \ll -\ell_{\alpha}^-$, such that the zero modes are properly described by their asymptotic behavior, Eqs. (\ref{asympsie}), (\ref{asymppsis}) and (\ref{asymppsir}). Obviously, our final result must not depend on our choice of $\zeta_{\text{ini}}$. Using Eqs.~(\ref{psiphi}), (\ref{asympsie}) and (\ref{asymppsis}), and imposing the condition
$c_{\sigma}=0$ we obtain the relation:
\begin{equation}
\label{constrainonc} 
c_{\sigma} e^{-\frac{\sigma^-}{v_0}\zeta_{\text{ini}}} =
\left(\psi_{\sigma}(\zeta_{\text{ini}})+\frac{\psi_E(\zeta_{\text{ini}})}
{\sigma^-}-\frac{\psi_{\phi}(\zeta_{\text{ini}}) v_0}{(\sigma^-
)^2}\right) = 0.
\end{equation}
From Eqs. (\ref{asympsie})-(\ref{asymppsir}) and the constraint (\ref{constrainonc}) we obtain
\begin{eqnarray}
\label{asymppsisdir}
\psi_{\sigma}(\zeta)&=&\psi_{\sigma}(\zeta_{\text{ini}})-
\frac{\psi_{\phi}(\zeta_{\text{ini}})}{\sigma^-}(\zeta-\zeta_{\text{ini}}), \\
\label{asymppsirdir}
\psi_{\rho}(\zeta)&=&\psi_{\rho}(\zeta_{\text{ini}})+\frac{\psi_{\phi}
(\zeta_{\text{ini}})}{\sigma^-}(\zeta-\zeta_{\text{ini}}), \\
\label{asymppsiedir} \psi_E(\zeta)&=& -\psi_{\sigma}(\zeta_{\text{ini}}) \sigma^- +\psi_{\phi}(\zeta_{\text{ini}})(\zeta- \zeta_{\text{ini}} +\frac{v_0}{\sigma^-}).
\end{eqnarray}
From Eqs. (\ref{asymppsisdir})-(\ref{asymppsiedir}) one can easily see that if $\psi_{\phi}(\zeta_{\text{ini}})$ is nonvanishing, the zero modes diverge linearly with $\zeta$ as $\zeta \to -\infty$. Since at least one of the three required independent relevant modes must have $\psi_{\phi}(\zeta_{\text{ini}}) \neq 0$, we must include the computation of such a diverging mode in our analysis. For computational purposes, it is convenient to choose the three independent orthogonal sets of initial conditions at $\zeta = \zeta_{\text{ini}}$ so that only one of them contains a nonvanishing value of $\psi_{\phi}(\zeta_{\text{ini}})$. It can easily be shown that choosing another set of three {\it independent} initial conditions, lead to identical results for the MBC. With our convenient choice we can immediately extract from Eqs.~(\ref{asymppsisdir})-(\ref{asymppsiedir}) the initial conditions at $\zeta = \zeta_{\text{ini}}$ (taking Eq.~(\ref{constrainonc}) into account) for the two nondiverging modes $[\psi^{(i)}_{\sigma},\psi^{(i)}_{\rho},\psi^{(i)}_{E},\psi^{(i)}_{\phi}]$, where $i=1,2$:
\begin{equation}
\label{initial1}
\left(\begin{array}{c}\psi_{\sigma}^{(1)}(\zeta_{\text{ini}})\\
\psi_{\rho}^{(1)}(\zeta_{\text{ini}})\\ \psi_E^{(1)}(\zeta_{\text{ini}})\\
\psi_{\phi}^{(1)}(\zeta_{\text{ini}})\end{array}\right)=\left(\begin{array}{c}
0\\1\\ 0\\ 0\end{array}\right) \ ,
\end{equation}
and
\begin{equation}
\label{initial2}
\left(\begin{array}{c}\psi_{\sigma}^{(2)}(\zeta_{\text{ini}})\\
\psi_{\rho}^{(2)}(\zeta_{\text{ini}})\\ \psi_E^{(2)}(\zeta_{\text{ini}})\\
\psi_{\phi}^{(2)}(\zeta_{\text{ini}})\end{array}\right)=\left(\begin{array}{c}
1\\0\\ -\sigma^-\\ 0\end{array}\right) \ ,
\end{equation}
where the linearity of Eqs. (\ref{eq0mode1b})-(\ref{eq0mode4b}) was used to scale one of the coefficients to unity in each initial condition (\ref{initial1}) and (\ref{initial2}). The last set of initial conditions, for the linearly diverging zero mode $[\psi^{(3)}_{\sigma},\psi^{(3)}_{\rho},\psi^{(3)}_{E},\psi^{(3)}_{\phi}]$, is determined by requiring it to be orthogonal to the two vectors (\ref{initial1}) and (\ref{initial2}) (and to satisfy the constraint (\ref{constrainonc})). This condition gives, up to a normalization factor:
\begin{equation}
\label{initial3}
\left(\begin{array}{c}\psi_{\sigma}^{(3)}(\zeta_{\text{ini}})\\
\psi_{\rho}^{(3)}(\zeta_{\text{ini}})\\ \psi_E^{(3)}(\zeta_{\text{ini}})\\
\psi_{\phi}^{(3)}(\zeta_{\text{ini}})\end{array}\right)=\left(\begin{array}{c}
\sigma^-\\
0\\ 1\\ \frac{\sigma^-}{v_0}(1+(\sigma^-)^2)\end{array}\right) \ .
\end{equation}
Thus, in order to obtain the three zero modes relevant for the MBC equations, from Eq.~(\ref{BC2}), we have to solve Eqs.~(\ref{eq0mode1b})-(\ref{eq0mode4b}) in the regime $\zeta \in (-\infty,0)$ under the initial conditions, Eqs.~(\ref{initial1})-(\ref{initial3}). Computation of the linearly diverging mode $[\psi^{(3)}_{\sigma},\psi^{(3)}_{\rho},\psi^{(3)}_{E},\psi^{(3)}_{\phi}]$ and extracting nondiverging MBC equations from Eq.~(\ref{BC2}) corresponding to this mode involves nontrivial technical complications, which for the sake of clarity we defer to the following section. We first discuss the two MBC equations corresponding to the two converging zero modes
$[\psi^{(i)}_{\sigma},\psi^{(i)}_{\rho},\psi^{(i)}_{E},\psi^{(i)}_{\phi}]$, $i=1,2$.

\subsubsection{The nondiverging zero modes}
\label{sec:nondiv}

In this subsection we derive a general form of two MBC equations. The exact expressions for the coefficients are deferred to section~\ref{sec:exact}. As we show below, the general form of the two MBC equations we derive here is sufficient to deduce an important property: The perturbed electrostatic field $E_1(-\infty)$ does not vanish in the inner region behind the front. This observation emphasizes the necessity to derive the third MBC equation with the aid of the linearly diverging zero mode.

For the two nondiverging modes we can take $\zeta_{\text{ini}} = \zeta_{\text{c}} \to -\infty$. Substituting Eqs.~(\ref{initial1}) and (\ref{initial2}) in Eq.~(\ref{BC2}) we obtain the two
MBC equations:
\begin{equation}
\label{BCtemp1}
v_0\sigma_1(-\infty)=[A^{(1)}+\psi_E^{(1)}(0))]v_1-B^{(1)} \, \lambda \ ,
\end{equation}
and
\begin{equation}
\label{BCtemp2} v_0\sigma_1(-\infty)+\sigma^- \, E_1(-
\infty)=[A^{(2)}+\psi_E^{(2)}(0)]v_1-B^{(2)}\, \lambda \ .
\end{equation}
Let us reconsider now our perturbative expansion (\ref{expansion}). As we emphasized in previous sections, the only physical source of perturbations to the planar front solution in our analysis are the transverse perturbations which correspond to nonvanishing $\kappa = \lambda \epsilon$. Thus, we formally write the perturbation to front velocity as $\epsilon\, v_1$, requiring it to vanish for $\epsilon =0$. The calculations presented so far, however, do not involve any assumption on the actual value of $v_1$, except $\epsilon\, v_1 \ll v_0$. By keeping $\kappa=0$ in our former analysis, we can thus consider the formal limit process: $\epsilon \to 0$ with $\epsilon\, v_1$ small (compared to $v_0$), but finite. This simply means that we consider a perturbation of the electric field for a {\it pure planar front} which implies $E_1(-\infty)=0$. Repeating our former analysis with $\kappa=0$ in Eq.~(\ref{laplacian}) we observe that the inhomogeneous term in Eq.~(\ref{expansion3}) vanishes, leading to $B=0$ in Eq. (\ref{eqB}). Substituting $B=0$ and $E_1(-\infty)=0$ in Eqs. (\ref{BCtemp1}) and (\ref{BCtemp2}) we obtain the nontrivial identity:
\begin{equation}
\label{BCtemp1b} v_0\sigma_1(-\infty)=[A^{(1)}+\psi_E^{(1)}(0)]v_1 =
[A^{(2)}+\psi_E^{(2)}(0)]v_1 \ ,
\end{equation}
which implies
\begin{equation}
\label{consistencycheck}
A^{(1)}+\psi_E^{(1)}(0)= A^{(2)}+\psi_E^{(2)}(0) \ .
\end{equation}
This important identity will be confirmed in Sec.~\ref{sec:mbc}
after finding the exact solution for the converging zero modes, and evaluating
the
actual values of $A^{(i)}+\psi_E^{(i)}(0)$, for $i=1,2$.

Furthermore, substituting the identity (\ref{consistencycheck}) in
Eqs. (\ref{BCtemp1}) and (\ref{BCtemp2}) we obtain the equation:
\begin{equation}
\label{e1inf}
- E_1(-\infty)=\frac{B^{(2)}-B^{(1)}}{\sigma^-}\,
\lambda \ .
\end{equation}
This relation implies that the electric field behind our curved fronts does not vanish if $B^{(1)}\ne
B^{(2)}$.

Finally, let us use the relation (\ref{e1inf}) to derive a correction to the boundary condition
(\ref{GLkatze}). Integrating the equation $\partial_{\zeta}\phi_1=-E_1$, we get:
\begin{equation} \label{integratephi}
\phi_1(\zeta_{\text{c}})-\phi_1(0)=\int_{\zeta_{\text{c}}}^0 d\zeta\,
E_1(\zeta).
\end{equation}
Since Eq. (\ref{e1inf}) implies that $E_1(\zeta)$ reaches a constant value $E_1(-\infty)$ as $\zeta\to - \infty$, substituting this in Eq.~(\ref{integratephi}) we obtain the asymptotic form of
$\phi_1(\zeta)$:
\begin{eqnarray}
\label{expphi1}
\phi_1(\zeta_{\text{c}})-\phi_1(0)&=&\int_{\zeta_{\text{c}}}^b
d\zeta\,
E_1(\zeta)+\int_{b}^0 d\zeta\, E_1(\zeta), \nonumber \\
&=&-\zeta_{\text{c}}E_1(-\infty)+W_1 \ ,
\end{eqnarray}
where $W_1$ is a constant. This relation will be used in Sec.~\ref{sec:newMBC}.

\subsection{Exact solutions for the zero modes}
\label{sec:exact}

In this section we find exact solutions of Eqs.~(\ref{eq0mode1b})-(\ref{eq0mode4b}) under the three sets of initial conditions, Eqs.~(\ref{initial1})-(\ref{initial3}), thus obtaining the three zero modes relevant for the MBC equations derived from Eq.~(\ref{BC2}). The calculations are lengthy and many technical details are delegated to appendix \ref{appendix2}. Since the final expressions we obtain are considerably complicated, we support our analytic calculations by comparison to numerical solutions of the system of ODEs (\ref{eq0mode1b})-(\ref{eq0mode4b}). To avoid cumbersome notations, we omit the superscript distinguishing between different zero modes in general expressions that apply for all modes.

\subsubsection{Analytic solution}
\label{analyticsol}

Our starting point is Eqs. (\ref{eq0mode1b})-(\ref{eq0mode4b}). With some algebraic manipulations (see appendix \ref{appendix2}), these equations can be brought to the form:
\begin{eqnarray}
\label{psis+psirsol2}
&&-
(v_0+E_0)\partial_{\zeta}^2(\psi_{\sigma}+\psi_{\rho})+f_0\partial_{\zeta}(\psi_{\sigma}+\psi_{\rho})\nonumber \\&&  + \frac{c_{\phi}E_0}{v_0}=0 \ , \\
\label{psisigmasol2}
&&E_0\partial_{\zeta}\psi_{\sigma}=-v_0\partial_{\zeta}(\psi_{\sigma}+\psi_{\rho})+f_0(\psi_{\sigma}+\psi_{\rho}) \ , \\
\label{psiesol}
&&\partial_{\zeta}\psi_E = \sigma_0 f_0'(\psi_{\sigma}+\psi_{\rho})- (\partial_{\zeta}\sigma_0)\psi_{\sigma}
+ c_{\phi} \ .
\end{eqnarray}
Notice the structure of this set of equations: Eq.~(\ref{psis+psirsol2}) is an inhomogeneous first order ODE for $\partial_{\zeta}(\psi_{\sigma}+\psi_{\rho})$ and can thus be solved analytically. Once this solution is substituted in Eq.~(\ref{psisigmasol2}), this last equation becomes also an inhomogeneous first order ODE for $\psi_{\sigma}$, and is thus also analytically solvable. Finally, substitution of the solutions to Eqs.~(\ref{psis+psirsol2}) and (\ref{psisigmasol2}) yields an inhomogeneous first order ODE for $\psi_E$, which is again analytically solvable. In the rest of this section we follow this procedure to obtain general analytic expressions with appropriate constants of integrations, which are then determined by imposing the initial conditions, Eqs.~(\ref{initial1})-(\ref{initial3}).

The general solution of Eq. (\ref{psis+psirsol2}) has the form:
\begin{equation}
\partial_{\zeta}(\psi_{\sigma}+\psi_{\rho})(\zeta)=C(\zeta)e^{g(\zeta)} \ ; \ g(\zeta)=\int_{-\infty}^{\zeta} dx
\frac{f_0(x)}{v_0+E_0(x)},
\end{equation}
where the expression of $C(\zeta)$ is given by (\ref{cczeta}). Integrating from $\zeta=-\infty$ to $\zeta$ (the sum $\psi_{\sigma}+\psi_{\rho}$ does not diverge at $\zeta=-\infty$, see Eqs. (\ref{asymppsisdir}) and (\ref{asymppsirdir})) and using the change of variables $y=E_0(x)$ to calculate the resulting integrals, we obtain the final form
\begin{equation}
\label{psis+psir2} (\psi_{\sigma}+\psi_{\rho})(\zeta)=c_{\sigma
\rho}+\frac{c_{\phi}}{v_0}\int_{0}^{E_0(\zeta)}dx\,
\frac{(v_0+x)}{\rho(x)^2} \ ,
\end{equation}
where
\begin{equation}
    \label{csigmarho}
    c_{\sigma \rho}=(\psi_{\sigma}+\psi_{\rho})(-\infty).
\end{equation}
The explicit steps leading to this expression are given in appendix \ref{appendix2b}.

Substituting Eq. (\ref{psis+psir2}) in Eq.~(\ref{psisigmasol2}), we obtain after integration and using the same variable transformation $y=E_0(x)$, the complicated expression (\ref{psis}) given in appendix \ref{appendix2c}. This expression can be written in the following compact form
\begin{equation}
\label{psisshort}
\psi_{\sigma}(\zeta) = \frac{c_{\phi}}{\sigma^-} \zeta_{\text{ini}} +
a_{\sigma}(\zeta) + b_{\sigma}(\zeta) e^{\lambda \zeta_{\text{ini}}}\ ,
\end{equation}
where $a_{\sigma}(\zeta)$ is given by Eq.~(\ref{express-asig}) and $b_{\sigma}(\zeta)$ is a regular function of $\zeta$ whose exact form does not affect our analysis since we will consider later the limit $|\zeta_{\text{ini}}|\gg 1$ and the last term of Eq.~(\ref{psisshort}) will be as small as we want. Notice that if $c_{\phi} \neq 0$, as is the case for the linearly diverging mode $[\psi^{(3)}_{\sigma},\psi^{(3)}_{\rho},\psi^{(3)}_{E},\psi^{(3)}_{\phi}]$, the first term on the RHS of Eq.~(\ref{psisshort}) diverges as the auxiliary parameter $\zeta_{\text{ini}} \to -\infty$. As we show below, this singular dependence on $\zeta_{\text{ini}}$ cancels out and does not appear in the final form of the MBC equation extracted from Eq.~(\ref{BC2}) corresponding to this mode. Since in deriving Eq.~(\ref{BC2}) in Sec.~\ref{zeromodesmbc} we assumed, however, that $\psi_{\sigma}(0)$ is finite, we must carry the $\zeta_{\text{ini}}$-dependent terms that appear in Eq.~(\ref{psisshort}) and other related expressions throughout our analysis.

Substituting Eqs.~(\ref{psis+psir2}) and (\ref{psisshort}) (whose full form is given by Eq.~(\ref{psis})) in Eq.~(\ref{psiesol}), integrating and using again the variable transformation $y=E_0(x)$ we obtain the cumbersome expression (\ref{psie}) given in appendix~\ref{appendix2d} (notice that to simplify this expression, we used for the first time the explicit form (\ref{alphaTown}) of $\alpha(x)$). This expression can also be written in the following compact form similarly to Eq.~(\ref{psisshort})
\begin{equation}
\label{psieshort}
\psi_{E}(\zeta) = -\frac{c_{\phi}}{\sigma^-} \sigma_0(\zeta) \zeta_{\text{ini}}+a_{E}(\zeta) + b_{E}(\zeta) e^{\lambda \zeta_{\text{ini}}}\ ,
\end{equation}
where $a_{E}(\zeta)$ is given by Eq.~(\ref{express-aE}) and $b_{E}(\zeta)$ is a regular function of $\zeta$ whose exact form does not affect our analysis.

\subsubsection{Comparison with numerical results}
\label{comp-with-num}

\begin{figure}[!hbtp]
\centerline{\includegraphics[width=\columnwidth,clip]{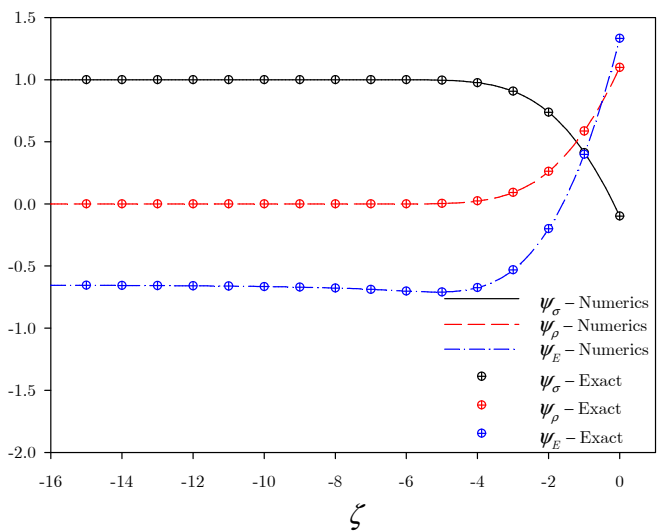}}
\caption{Comparison between numerical solution of Eqs.~(\ref{eq0mode1b})-
(\ref{eq0mode4b}) and the exact solution reported in Sec.~\ref{sec:exact} for
the initial condition (\ref{initial2}), with $E_{\infty}=-2$ and
$\zeta_{\text{ini}}=-40$.} \label{fig03}
\end{figure}

\begin{figure}[!hbtp]
\centerline{\includegraphics[width=\columnwidth,clip]{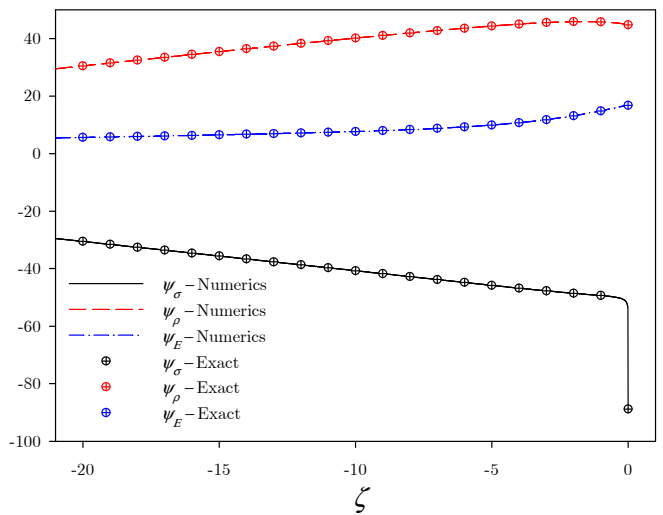}}
\caption{Comparison between numerical solution of Eqs.~(\ref{eq0mode1b})-
(\ref{eq0mode4b}) and the exact solution reported in Sec.~\ref{sec:exact} for
the initial condition (\ref{initial3}), with $E_{\infty}=-1$ and
$\zeta_{\text{ini}}=-50$.} \label{fig04}
\end{figure}

To check the analytic formulas (\ref{psis+psir2}), (\ref{psisshort}) and (\ref{psieshort}) with the
complicated expressions, Eqs.~(\ref{psis+psir}), (\ref{psis}) and (\ref{psie}), we used MATHEMATICA to obtain numerical solutions of the zero modes Eqs.~(\ref{eq0mode1b})-(\ref{eq0mode4b}) under the
initial conditions (\ref{initial1})-(\ref{initial3}), for few representative values of $\zeta_{\text{ini}}$. In Fig.~\ref{fig03}, we present the numerical solution obtained with
the initial conditions for one of the two nondiverging modes, Eq.~(\ref{initial2}), evaluated at $\zeta_{\text{ini}}=-40$ for $E_{\infty}=-2$.
In Fig.~\ref{fig04}, we report the numerical solution for the linearly diverging mode, obtained with the initial conditions, Eq.~(\ref{initial3}), evaluated at $\zeta_{\text{ini}}=-50$ for $E_{\infty}=-1$. Figs.~\ref{fig03} and \ref{fig04} exhibit a excellent agreement between our analytic expressions and the numerical solutions.

\subsubsection{General expression for the relevant coefficients}

In this section, we use the analytic solutions derived in Sec.~\ref{analyticsol} to provide compact expressions for the coefficients $A$, $A+\psi_E(0)$ and $B$, required for deriving the MBC equations from Eq.~(\ref{BC2}). The complete forms of these expressions are given in appendix~\ref{appendix2}. Using the definition of $A$, Eq.~(\ref{eqA}), and the exact expressions of $\psi_{\sigma}+\psi_{\rho}$, Eq.~(\ref{psis+psir2}), and of $\psi_{\sigma}$, Eq.~(\ref{psisshort}) (given in its full form in Eq.~(\ref{psis})), we obtain the explicit form for $A$, Eq.~(\ref{expressionA}) (see apeendix~\ref{app:expressA}). This expression can be written in the compact form
\begin{equation}
    \label{expressAshort}
    A=\frac{\sigma_0(0)c_{\phi}}{\sigma^-} \zeta_{\text{ini}}+ a_A+b_A
e^{\lambda \zeta_{\text{ini}}},
\end{equation}
where $a_{A}$ is given by Eq.~(\ref{express-aA}) and the exact form of $b_{A}$ is not required for our analysis. The exact form of the coefficient $A+\psi_E(0)$ is given in Eq.~(\ref{A+psie0}) which is obtained using the expressions (\ref{expressAshort}) and (\ref{psieshort}) for $A$ and $\psi_E$
respectively (see Eqs.~(\ref{psie}) and (\ref{expressionA}) for their complete expressions). The compact form for $A+\psi_E(0)$ is given by
\begin{equation}
    \label{expressA+psieshort}
    A+\psi_E(0)=a_A+a_E(0)+b_E(0) e^{\lambda \zeta_{\text{ini}}}.
\end{equation}
Notice that this last expression does not diverge when $\zeta_{\text{ini}} \to -\infty$. The term $a_A+a_E(0)$ is given by Eq.~(\ref{aA+aE}). Using the definition of $B$, Eq.~(\ref{eqB}), and the expression (\ref{psieshort}) for $\psi_E(\zeta)$, we obtain the complicated form of $B$, Eq.~(\ref{expressionB}). The compact form for $B$ is given by
\begin{equation}
    \label{expressBshort}
    B=\frac{c_{\phi}v_0^2}{\sigma^-} \zeta_{\text{ini}}+ a_B+b_B e^{\lambda
\zeta_{\text{ini}}},
\end{equation}
where $a_{B}$ is given by Eq.~(\ref{express-aB}) and the exact form of $b_{B}$ is again not required our analysis.

\subsection{Completing inner region analysis}
\label{sec:mbc}

We here complete the inner region analysis by returning to Eq.~(\ref{BC2}) and using the results obtained in Sec.~\ref{sec:exact} to derive MBC equations for the curved front.

First, let us rewrite Eq.~(\ref{BC2}) in the form
\begin{eqnarray}
    \label{BC3}
    &-&(A+\psi_E(0))v_1+B \lambda=E_1(\zeta_{\text{c}})\left(-\psi_{\sigma}(\zeta_{\text{ini}}) \sigma^-
    \right. \nonumber \\  &+& \left. c_{\phi}(\zeta_{\text{c}}- \zeta_{\text{ini}} +\frac{v_0}{\sigma^-})\right)  + c_{\phi}(\phi_1(\zeta_{\text{c}})-\phi_1(0))\nonumber \\
&-&c_{\sigma \rho}v_0\sigma_1(\zeta_{\text{c}}),
\end{eqnarray}
where we used Eqs.~(\ref{psiphi}), (\ref{asymppsiedir}) and (\ref{csigmarho}) (taking already the limit
$\zeta_{\text{c}} \to -\infty$). To further facilitate our analysis, let us first write the expressions of the
coefficients $A+\psi_E(0)$ and $B$ in the following form
\begin{eqnarray}
    \label{expressionAb}
    A+\psi_E(0)&=&\gamma_1 c_{\phi}+\gamma_2 c_{\sigma \rho}+ \gamma_3
\psi_{\sigma}(\zeta_{\text{ini}}) \nonumber \\ &+&\gamma_4
\psi_E(\zeta_{\text{ini}}) +b_E(0) e^{\lambda \zeta_{\text{ini}}},  \\
    \label{expressionBb}
    B&=&(\frac{v_0^2}{\sigma^-}\zeta_{\text{ini}}+\beta_1) c_{\phi}+\beta_2
c_{\sigma \rho}+ \beta_3 \psi_{\sigma}(\zeta_{\text{ini}}) \nonumber \\
&+&\beta_4 \psi_E(\zeta_{\text{ini}})+b_B e^{\lambda \zeta_{\text{ini}}}.
\end{eqnarray}
The coefficients $\gamma_j$ and $\beta_j$, $j=1,\cdots,4$, depend only on $E^+$
and can be easily extracted from Eqs.~(\ref{aA+aE}) and (\ref{express-aB}):
\begin{eqnarray}
\label{gamma1}
\gamma_1&=&\int_0^{E^+}dx\, [\alpha(x)-\alpha(E^+)]\frac{(v_0-x)}{\rho(x)^2} \\
\label{gamma2}
\gamma_2&=&v_0\alpha(E^+) \\
\label{gamma3}
\gamma_3&=&\sigma^-\\
\label{gamma4}
\gamma_4&=&1.
\end{eqnarray}
\begin{eqnarray}
\label{beta1}
\beta_1&=& v_0\int_0^{E^+} dx \frac{(v_0-x)\alpha(x)}{x\rho(x)}\int_0^{x}dy
\frac{(v_0-y)}{\rho(y)^2}  \\
&+&\frac{1}{\sigma^-}\int_0^{E^+} dx \frac{(v_0^2-
x^2)}{x\rho(x)^2}\left[\rho(x)v_0-\sigma^-(v_0-x) \right] \nonumber \\
\label{beta2}
\beta_2&=&-v_0^2\int_{0}^{E^+}dx \frac{(v_0-x)\alpha(x)}{x\rho(x)} \\
\label{beta3}
\beta_3&=&v_0^2-\sigma^-\int_{0}^{E^+}dx \frac{v_0-x}{\rho(x)}\\
\label{beta4}
\beta_4&=&-\int_{0}^{E^+}dx \frac{v_0-x}{\rho(x)}.
\end{eqnarray}
The contributions of last the term on the RHS of Eqs.~(\ref{expressionAb}) and (\ref{expressionBb}) can be made arbitrarily small by choosing $|\zeta_{\text{ini}}|$ sufficiently large and can thus be neglected in the rest of our analysis.

The first MBC equation is obtained by using the values of the nondiverging zero mode corresponding to the initial condition (\ref{initial1}) in Eqs.~(\ref{BC3}), (\ref{expressionAb}) and (\ref{expressionBb}). We obtain
\begin{equation}
    \label{BCfinal1}
    v_0\sigma_1(-\infty)=\gamma_2 v_1- \beta_2 \lambda,
\end{equation}
where we took the limit $\zeta_{\text{c}} \to -\infty$. The second MBC equation is obtained in a similar way by using the values of the second nondiverging zero mode corresponding to the initial condition (\ref{initial2}) in Eqs.~(\ref{BC3}), (\ref{expressionAb}) and (\ref{expressionBb}). Using Eq.~(\ref{BCfinal1}), we obtain
\begin{equation}
    \label{BCfinal2}
    -E_1(-\infty)=\frac{v_0^2}{\sigma^-} \lambda.
\end{equation}
As was pointed out already in Eq.~(\ref{e1inf}), this last equation implies that the electrostatic field does not vanish behind the front. Notice also that, as mentioned in the discussion following Eq.~(\ref{consistencycheck}), it is now easy to verify explicitly the identity
\begin{equation}
    \label{consistencycheck2}
    A^{(1)}+\psi_E^{(1)}(0)=A^{(2)}+\psi_E^{(2)}(0)=\gamma_2.
\end{equation}

Using the nondiverging zero modes to extract MBC equations (\ref{BCfinal1}) and (\ref{BCfinal2}) from (\ref{BC3}) was relatively straightforward since all terms are finite at both end points $\zeta=0$ and $\zeta=\zeta_{\text{c}} \to -\infty$. Naively, one may anticipate that since $\psi_E^{(3)}(\zeta_{\text{c}})\sim \zeta_{\text{c}}$ it is impossible to extract physically meaningful MBC equation corresponding to this mode from (\ref{BC3}). We show however below that taking appropriately the asymptotic limits $\zeta_{\text{c}} \to -\infty$ and the auxiliary parameter
$|\zeta_{\text{ini}}| \gg 1$ leads to cancellation of the diverging terms and to an additional nontrivial MBC equation.

In order to derive this third MBC equation, using the initial condition (\ref{initial3}), we first substitute (\ref{expphi1}) in Eq.~(\ref{BC3}) and use Eqs.~(\ref{expressionAb}) and (\ref{expressionBb}) to obtain
\begin{eqnarray}
\label{xxx}
&-&\left[\gamma_1 c_{\phi}+\sigma^-(\gamma_2+ \gamma_3) +\gamma_4 \right]v_1
\nonumber \\ &+&\left[(\frac{v_0^2}{\sigma^-}\zeta_{\text{ini}}+\beta_1)
c_{\phi}+\sigma^-(\beta_2 + \beta_3) +\beta_4 \right] \lambda  \nonumber \\
&=&c_{\phi} W_1-\sigma^- v_0\sigma_1(\zeta_{\text{c}}) 
+E_1(\zeta_{\text{c}})\left[ c_{\phi}v_0/\sigma^- -(\sigma^-)^2\right] \nonumber \\ &-&c_{\phi}\zeta_{\text{ini}}E_1(\zeta_{\text{c}}),
\end{eqnarray}
where $c_\phi$ is given by (\ref{initial3}) (using Eq.~(\ref{psiphi})). 
Notice that the divergence of $\psi_E^{(3)}$ and $\phi_1$ with $\zeta_{\text{c}}$ cancels out provided $|\zeta_{\text{c}}|$ is large enough such as $E_1(\zeta_{\text{c}})$ has reached its plateau value. We can now take the limit $\zeta_{\text{c}}\to -\infty$ such as the limits in Eqs.~(\ref{BCfinal1}) and (\ref{BCfinal2}) are reached and can be used in Eq.~(\ref{xxx}). We then also notice that the two terms proportional to $\zeta_{\text{ini}}$ in Eq.~(\ref{xxx})
cancel. After some manipulations, our third MBC equation can be written as
\begin{equation}
    \label{BCfinal3}
    W_1=- \omega_1 \, v_1 + \omega_2 \lambda
\end{equation}
where
\begin{eqnarray}
    \label{BCfinal3a}
    \omega_1&=&\gamma_1+\frac{v_0}{\sigma^-} \\
\label{BCfinal3b}   \omega_2&=&\beta_1-\frac{v_0}{\sigma^-
}\int_0^{E^+}dx\frac{v_0-x}{\rho(x)}+\frac{v_0^3}{(\sigma^-)^2}
\end{eqnarray}

\begin{figure}[!hbtp]
\centerline{\includegraphics[width=\columnwidth, clip]{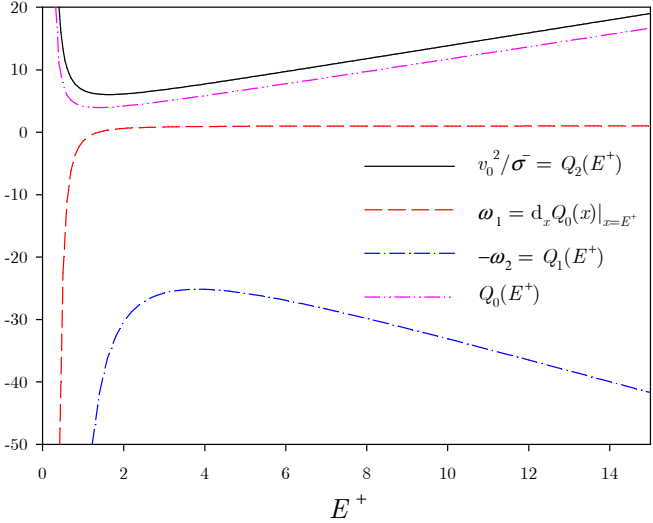}}
\caption{$v_0^2/\sigma^-=Q_2(E^+)$, $\omega_1$, $\omega_2=-Q_1(E^+)$ and
$Q_0(E^+)$ as a function of $v_0=E^+$. See Sec.~\ref{sec1} for the relation
between the coefficients $Q_i$ and the MBC equations.} \label{fig05}
\end{figure}

\begin{figure}[!hbtp]
\centerline{\includegraphics[width=\columnwidth, clip]{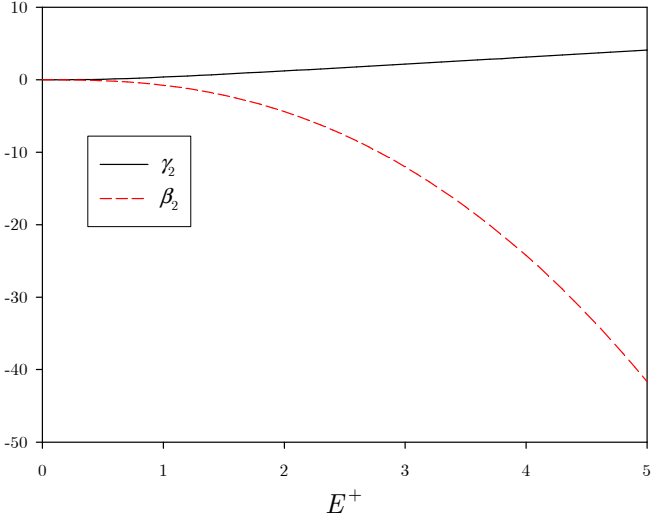}}
\caption{The coefficients $\gamma_2(E^+)$ and $\beta_2(E^+)$ which
determine the modified electron and ion density behind the front
(see Eq. (\ref{BCfinal1}))} \label{fig06}
\end{figure}

In Fig.~\ref{fig05}, we present the dependence of the coefficients in Eqs.~(\ref{defq0}), (\ref{BCfinal2}) and (\ref{BCfinal3}) on the variable $E^+$. The relations between these functions and the functions $Q_i$, used in Eqs.~(\ref{BCphi-}) and (\ref{BCpotjump}), are derived later in this section. In Fig.~\ref{fig06}, we present the dependence of the coefficients in Eq.~(\ref{BCfinal1}) on the variable $E^+$.

Notice the structure of the MBC equations (\ref{BCfinal1}), (\ref{BCfinal2}) and (\ref{BCfinal3}) derived above. As we will show below, the last two equations give rise to the MBC equations (\ref{BCphi-}) and (\ref{BCpotjump}), whereas the first one determines the excess density of ionized matter behind the curved front which is thus ``slaved'' by the dynamics specified by Eqs.~(\ref{laplaceahead})-(\ref{velocityfront}).

Let us discuss Eqs.~(\ref{BCfinal1}), (\ref{BCfinal2}) and (\ref{BCfinal3}), in view of results obtained in Ref.~\cite{Man04} for planar front solution. For purely planar front, the only possible perturbation of the quantities ($\sigma_0$, $\rho_0$ and $E_0$) are due to modification of the electric field ahead of the front $E_0(0)=E_{\infty}$. As already discussed in Sec.~\ref{sec:nondiv}, in our formalism this is equivalent to keeping in Eqs.~(\ref{BCfinal1}) and (\ref{BCfinal3}) only the terms proportional to $v_1$. In this case we also have $E_1(-\infty)=0$, which implies that Eq.~(\ref{expphi1}) reduces to
\begin{equation}
	\label{expphi1b}
	\phi_1(-\infty)-\phi_1(0)=W_1.
\end{equation}

A first test concerns our Eq.~(\ref{BCfinal1}). In Ref.~\cite{Man04} the following relation was derived
\begin{equation}
    \label{xxx1}
    -\frac{d\sigma^-}{dE^+}\equiv \frac{d\sigma_0(-\infty)}{dE_0(0)}=-\alpha(E^+),
\end{equation}
with $E_0(0)=-E^+$. This relation actually follows directly from the definition of $\sigma^-$ (\ref{edensminfty}). Since for infinitesimal perturbations $d\sigma_0(-\infty)=\sigma_1(-\infty)$ and $dE_0(0)=E_1(0)$ we obtain from Eq.~(\ref{xxx1})
\begin{equation}
    \label{check1}
    \sigma_1(-\infty)=-\alpha(E^+)\, E_1(0)\equiv \alpha(E^+)\, v_1.
\end{equation}
Keeping on the RHS of Eq.~(\ref{BCfinal1}) only terms proportional to $v_1$ and using (\ref{gamma2}) we recover Eq.~(\ref{check1}).

The second test stems from comparing our MBC Eq.~(\ref{BCfinal3}) to the planar front relation Eq.~(\ref{defq0}), which for infinitesimal perturbations, $\phi_1$ and $E_1$, recast the form
\begin{equation}
    \label{check2a}
    \phi_1(-\infty)-\phi_1(0)=-\frac{dQ_0(E^+)}{dE^+}\, v_1.
\end{equation}
A simple computation leads to
\begin{equation}
    \label{check2b}
    \frac{dQ_0(E^+)}{dE^+}=\frac{1}{\alpha(E^+)}+\int_0^{E^+}\frac{dx}{\rho(x)
}-\alpha(E^+)\int_0^{E^+}dx\frac{E^+ -x}{\rho(x)^2}.
\end{equation}
Keeping again only terms proportional to $v_1$ on the RHS of Eq.~(\ref{BCfinal3}), using Eq.~(\ref{expphi1b}) and comparing with (\ref{check2a}) implies that the following equality
\begin{equation}
    \label{relq0w1}
    \frac{dQ_0(E^+)}{dE^+}= \omega_1,
\end{equation}
must be satisfied. To check this we use the definition (\ref{BCfinal3a}) together with (\ref{gamma1}) to obtain
\begin{eqnarray}
    \label{xxx2}
    \omega_1&=&-\alpha(v_0)\int_0^{v_0}dx \frac{v_0-x}{\rho(x)^2} \nonumber \\
&+&\int_0^{v_0}dx \frac{\alpha(x)(v_0-x)}{\rho(x)^2}+\frac{v_0}{\sigma^-}
\end{eqnarray}
Using Eq.~(\ref{rho}), integrating by parts the second integral (using $d_x(1/\rho(x))=\alpha(x)/\rho(x)^2$) and taking the appropriate limit of boundary terms, Eq.~(\ref{xxx2}) becomes identical, as required, to Eq.~(\ref{check2b}). These last two tests demonstrate consistency of our calculation with the previous results of \cite{Man04}.

\section{A physical discussion of the curvature corrections}
\label{sec:refine}

\subsection{The new moving boundary conditions}
\label{sec:newMBC}

In this section, we derive a refined version of the boundary model which includes curvature correction. We first write the equations of the model and we discuss it below. It consists of Laplace equations for both outer regions
\begin{eqnarray}
\nabla^2 \phi &=& 0 \quad \text{in} \quad \Omega^+ \label{laplaceahead} \\
\nabla^2 \phi &=& 0 \quad \text{in} \quad \Omega^- \label{laplaceouter}
\end{eqnarray}
and the moving boundary conditions
\begin{eqnarray}
\hat{{\bf n}} \cdot \nabla \phi^- &=& Q_2(\hat{{\bf n}} \cdot \nabla \phi^+)
\kappa \label{BCphi-} \\
\phi^+-\phi^- &=& Q_0(\hat{{\bf n}} \cdot \nabla \phi^+) + Q_1(\hat{{\bf n}} \cdot \nabla \phi^+) \kappa \label{BCpotjump} \\
v_{n} &=& \hat{{\bf n}} \cdot \nabla\phi^+ \label{velocityfront},
\end{eqnarray}
where the coefficients $Q_i$ depend on the electrostatic field ahead the front, and are given by analytic formulas derived from the planar front solution as follows
\begin{eqnarray}
	\label{q0}
	Q_0(y)&=&\int_0^{y} dz\, \frac{y-z}{\rho(z)}, \\
	\label{q1}
	Q_1(y)&=&-\omega_2(y), \\
	\label{q2}
	Q_2(y)&=&\frac{y^2}{\sigma^-(y)}	
\end{eqnarray}
and where $\rho$, $\sigma^-$ and $\omega_2$ are given by (\ref{rho}), (\ref{edensminfty}) and (\ref{BCfinal3b}) respectively. 

From inner analysis performed in the previous sections, and as already discussed in Sec.~\ref{phi+}, we know that the electron and ion densities vanish precisely at the discontinuity line of the front part of the ionization front where the matching with the outer region $\Omega^+$ has to be done. Since there is no charge in $\Omega^+$ the only consistent equation for the electric potential in this region is given by (\ref{laplaceahead}). On the other hand, we know also that the space charge density vanishes at the back part of the inner region (see Eqs.~(\ref{rholimit}) and (\ref{sig1=rho1})). Consequently Eq.~(\ref{laplaceouter}) for the outer region $\Omega^-$ is also consistent with the inner analysis. Another possible equation for the electric potential in $\Omega^-$ is given by $\nabla^2 \phi = \sigma-\rho$ with the necessary boundary condition $\sigma-\rho=0$ on $\Gamma^-$ in order to be consistent with the inner analysis. However this equation introduces new unknowns, the electron and ion densities in $\Omega^-$, for which additional evolution equations would be needed. In Sec.~\ref{CCApprox}, we discuss the constant conductivity approximation introduced by Eq.~(\ref{laplaceouter}).

The boundary equation (\ref{BCphi-}) is obtained directly from Eq.~(\ref{BCfinal2}). Indeed, as $\epsilon \to 0$, the inner region reduces to the curve $\Gamma$ (the sharp interface) and 
\begin{equation}
	\label{eqfore1}
	E_1(-\infty)\equiv \hat{{\bf n}} \cdot \nabla \phi_1^-.
\end{equation}
On the other hand we know that
\begin{equation}
	\label{eqforv0}
	v_0\equiv E^+=\hat{{\bf n}} \cdot \nabla \phi^+
\end{equation}
Consequently we have in the first order in $\epsilon$ using the expansion (\ref{expansion}) and Eqs.~(\ref{BCphi-}), (\ref{q2}), (\ref{eqfore1}) and (\ref{eqforv0})
\begin{equation}
	\hat{{\bf n}} \cdot \nabla \phi^-=\hat{{\bf n}} \cdot \nabla (\phi_0^- +\epsilon \phi_1^-)=\frac{v_0^2}{\sigma^-(v_0)} \kappa.
\end{equation}
Since $\hat{{\bf n}} \cdot \nabla \phi_0^- =0$ and $\epsilon = \kappa / \lambda$, we just recover Eq.~(\ref{BCfinal2}).

The boundary equation (\ref{BCpotjump}) is obtained using Eqs.~(\ref{expphi1}), (\ref{BCfinal2}) and (\ref{BCfinal3}) which reads
\begin{equation}
	\phi_1(0)-\phi_1(\zeta_c)=\omega_1 v_1 - \left(\omega_2  + \frac{v_0^2}{\sigma^-(v_0)} \zeta_c\right)\lambda.
\end{equation}
The quantity $\zeta_c$ lies in the inner region and will coincide with $0^-$ when $\epsilon \to 0$ whereas $0 \to 0^+$. Using the expansion (\ref{expansion}) together with Eqs.~(\ref{defq0}), (\ref{v1}), (\ref{BCfinal3}), (\ref{relq0w1}), (\ref{q1}) and (\ref{eqforv0}) we obtain
\begin{eqnarray}
\phi^+ - \phi^- &=& \phi_0^+ - \phi_0^- +\epsilon (\phi_1^+ - \phi_1^-) \nonumber \\
     &=& Q_0(E^+)- \epsilon W_1 = Q_0(E^+)+ \epsilon (\omega_1 v_1- \omega_2 \lambda) \nonumber \\
     &=& Q_0(\hat{{\bf n}} \cdot \nabla \phi^+) + Q_1(\hat{{\bf n}} \cdot \nabla \phi^+) \kappa,
\end{eqnarray} 
which is just the boundary condition (\ref{BCpotjump}).

Eq.~(\ref{BCpotjump}) contains a curvature correction to the ``kinetic undercooling" boundary condition (\ref{GLkatze}). This correction might suppress non-regular solutions~\cite{Prok08} of the minimal regularized boundary model.

Eq.~(\ref{BCphi-}) has required us to take the finite conductivity of the ionized region into account (Eq.~(\ref{laplaceouter})); it expresses the field behind the ionization front in terms of the field ahead of it, of the curvature and of the conductivity in the ionized region.

\subsection{The field inside a propagating streamer}

Eq.~(\ref{BCphi-}) has an interesting direct physical consequence for streamers whose radius and field
enhancement evolve slowly during propagation within a comoving coordinate frame. Indeed, simulations of single streamers show that the field inside the streamer is constant in very good approximation (see for example Fig.~4 in \cite{brau07}). Such a constant interior field is a common ingredient of phenomenological streamer modeling. Here we can calculate this field: on the streamer axis we find
\begin{equation}
{\bf E}^-=-Q_2({\bf E}^+)\,\kappa\,\hat{\bf z}.
\end{equation}
That the field inside the streamer might approach some fixed constant value, is a common interpretation of experimental results. Here we have derived an explicit relation for this field; more precisely, it depends on the conductivity inside the streamer, on the front velocity and on the curvature. This explicit prediction will be discussed further and tested on simulations elsewhere~\cite{newFabian}.

\subsection{Large field limit}

Since the width of the front $\ell_{\alpha}^-(v_0)$, see Eqs.~(\ref{lambdaminus}) and (\ref{lambdaminus1}), is a monotonic decreasing function of $v_0$, we expect that the sharp interface will be a better approximation for large fields. Moreover, as seen in Fig.~\ref{fig05}, the coefficients $Q_i$ involved in the boundary conditions (\ref{BCphi-}) and (\ref{BCpotjump}) become linear in this regime which simplifies the model. In this section, we compute the asymptotic behaviors of the coefficients $Q_i$.  

From the definition (\ref{q0}) of $Q_0(v_0)$ we have
\begin{equation}
	\frac{d Q_0}{d v_0}=\lim_{x\to v_0} \frac{v_0-x}{\rho(x)}=\frac{1}{\alpha(v_0)}\sim 1 \quad \text{for}\quad v_0\to \infty.
\end{equation}
Consequently, we simply have
\begin{equation}
	\label{q0asymp}
	Q_0(v_0)\sim v_0 \quad \text{for}\quad v_0\to \infty.
\end{equation}
Using the same method, the limit of the constant conductivity behind the front (\ref{edensminfty}) takes the form
\begin{equation}
	\label{sigma-asymp}
	\sigma^-(v_0)\sim v_0 \quad \text{for}\quad v_0\to \infty,
\end{equation}
since
\begin{equation}
	\frac{d \sigma^-}{d v_0}=\alpha(v_0)\sim 1 \quad \text{for}\quad v_0\to \infty.
\end{equation}
We then deduce that the asymptotic behavior of $Q_2$ (\ref{q2}) is given by
\begin{equation}
	\label{q2asymp}
	Q_2(v_0)\sim v_0 \quad \text{for}\quad v_0\to \infty.
\end{equation}

The computation of the asymptotic behavior of $Q_1$ is a bit more lengthly. The expression of this coefficient is given by five terms. Let us compute the limit for each one and use the notation $q_i$ with $i=1,2, \cdots,5$ with $Q_1=\sum_{i=1}^5 q_i$. Using the results obtained above we easily find that
\begin{equation}
	q_1=-\frac{v_0^3}{(\sigma^-)^2}\sim -v_0 \quad \text{for}\quad v_0\to \infty.
\end{equation}
\begin{equation}
	q_2=\frac{v_0}{\sigma^-}Q_0(v_0)\sim v_0 \quad \text{for}\quad v_0\to \infty.
\end{equation}
The third term, actually contained in the expression of $\beta_1$ (\ref{beta1}), is given by
\begin{equation}
	q_3=-\frac{v_0}{\sigma^-}\int_0^{v_0}dx \frac{(v_0+x)(v_0-x)}{x \rho(x)}\equiv -\frac{v_0}{\sigma^-}\, \bar{q}_3(v_0).
\end{equation}
The factor in front of the integral tends to 1 for large $v_0$. The asymptotic behavior of the integral $\bar{q}_3$ is obtained as above:
\begin{equation}
	\frac{d \bar{q}_3}{d v_0}=2 \lim_{x\to v_0} \frac{v_0-x}{\rho(x)}=\frac{2}{\alpha(v_0)} \sim 2 \quad \text{for}\quad v_0\to \infty.
\end{equation}
We then find
\begin{equation}
	q_3 \sim -2 v_0 \quad \text{for}\quad v_0\to \infty.
\end{equation}
The fourth term is given by
\begin{equation}
	q_4 = \int_0^{v_0} dx \frac{(v_0+x)(v_0-x)^2}{x \rho(x)^2}.
\end{equation}
To obtain the asymptotic behavior we also use the derivative:
\begin{eqnarray}
	\frac{d q_4}{d v_0}&=&2\lim_{x\to v_0} \frac{(v_0-x)^2}{\rho(x)^2}=2\lim_{x\to v_0} \frac{(v_0-x)}{\rho(x) \alpha(x)} \nonumber \\
	 &=& \frac{2}{\alpha(v_0)^2} \sim 2 \quad \text{for}\quad v_0\to \infty.
\end{eqnarray}
We thus obtain
\begin{equation}
	q_4 \sim 2 v_0 \quad \text{for}\quad v_0\to \infty.
\end{equation}
Which means that $q_1+q_2+q_3+q_4 \sim {\cal O}((v_0)^0)$ for large $v_0$. Since we know from numerics that the asymptotic behavior of $Q_1$ is linear, this behavior is encoded in the last term:
\begin{equation}
	q_5=-v_0 \int_0^{v_0} dx \frac{(v_0-x)\alpha(x)}{x\rho(x)}\int_0^{x}dy \frac{(v_0-y)}{\rho(y)^2}.
\end{equation}
With the help of the following changes of variables $x=v_0 q$ and $y=v_0 p$ we get
\begin{equation}
	q_5=-v_0^4 \int_0^{1} dq \frac{(1-q)\alpha(v_0 q)}{q\rho(v_0 q)}\int_0^{q}dp \frac{(1-p)}{\rho(v_0 p)^2}.
\end{equation}
From the definition of $\rho(x)$ (\ref{rho}) we have
\begin{equation}
	\rho(v_0 z)=v_0\int_{z}^1 d\mu \, \alpha(v_0\mu),
\end{equation}
and from the definition of $\alpha(x)$ (\ref{alphaTown}) we know that
\begin{equation}
	\lim_{v_0\to \infty} \alpha(v_0 z)=\theta(z),
\end{equation}
where $\theta(z)$ is the Heaviside's function. We then get
\begin{equation}
	q_5\sim v_0 \int_0^1 dq \frac{\ln(1-q)}{q} 
	  = -\frac{\pi^2}{6} v_0 \quad \text{for}\quad v_0\to \infty.
\end{equation}
Finally the asymptotic behavior of $Q_1(v_0)$ is given by
\begin{equation}
	\label{q1asymp}
	Q_1(v_0)\sim -\frac{\pi^2}{6}v_0 \quad \text{for}\quad v_0\to \infty.
\end{equation}

For large electric fields ($E^+\gtrsim 4$), the boundary conditions (\ref{BCphi-})-(\ref{velocityfront}) reduce 
\begin{eqnarray}
\hat{{\bf n}} \cdot \nabla \phi^- &=&\kappa\, \hat{{\bf n}} \cdot \nabla \phi^+  \label{BCphi-asymp} \\
\phi^+-\phi^- &=& (1 - \pi^2 \kappa/6)\, \hat{{\bf n}} \cdot \nabla \phi^+  \label{BCpotjumpasymp} \\
v_{n} &=& \hat{{\bf n}} \cdot \nabla\phi^+ \label{velocityfrontasymp}.
\end{eqnarray}

\subsection{Algorithm}

Let us describe now an elementary algorithm for numerical solution of the front dynamics with the MBC Eqs.~(\ref{laplaceahead})-(\ref{velocityfront}). For simplicity let us consider a front characterized by a curve $(x_f(y,t),y)$, and far field boundary conditions: (i) $\phi=0$ at an electrode $(x=0,y)$ behind the moving front, and (ii) $\frac{\partial \phi}{\partial x} = -E_{\infty}$ at $x \approx L \gg \ell_{\alpha}^{-}$. For simplicity, let us further assume an initial state of the front $(x_f(y,0),y)$, where $x_f(y,0)=x_0+\delta f(y)$, $x_0 \gg \ell_{\alpha}^{-}$ and $|\delta f''(y)| \ll 1$ and where $\ell_{\alpha}^{-} \ll x_0 \ll L$. For such an initial condition, the initial expression for $Q_2$ in Eq.~(\ref{BCphi-}) is approximately $Q_2(|E_{\infty}|)$ (up to correction $O(\delta f''(y))$). The front dynamics
is then prescribed by the following numerical algorithm: 
\begin{enumerate}
\item solve Laplace Eq. (\ref{laplaceouter}) in the region $\Omega^-$
with boundary conditions (i) at $(x=0,y)$ and (\ref{BCphi-}) at
$(x\to (x_f(y,t))^-,y)$ .
\item Use the solution of step 1 to evaluate the potential $\phi^-(x\to
(x_f(y,t))^-,y)$.
\item Substitute the potential $\phi^{-}$ evaluated at step 2 in
Eq. (\ref{BCpotjump}), and solve the Laplace Eq.
(\ref{laplaceahead}) in the region $\Omega^+$ subject to the
boundary conditions (ii) and (\ref{BCpotjump}).
\item Use the solution of step 3 to evaluate the gradient of the potential
$\hat{{\bf n}} \cdot \nabla \phi^+(x \to x_f(y,t))^+,y)$ .
\item Advance the curve $(x_f(y,t),y)$ in a normal direction $\hat{{\bf n}}$
(towards $\Omega^+$) at a rate $\hat{{\bf n}} \cdot \nabla \phi^+$ found in step 4.
\item Modify the functions $Q_i$ according to Eqs.~(\ref{q0})-(\ref{q2}) in the MBC Eqs.~(\ref{BCphi-}) and (\ref{BCpotjump}) according to the value of $\hat{{\bf n}} \cdot \nabla
\phi^+$ found in step 4 and return to step 1.
\end{enumerate}

\section{The constant conductivity approximation in the ionized interior}
\label{CCApprox}

\subsection{The degree of ionization behind a front}

Eqs.~(\ref{edensminfty}), (\ref{rholimit}) show that the degree of ionization
$\sigma^-=\rho^-=\int_0^{E^+}d\mu\, \alpha(\mu)$ behind a uniformly translating front depends on the
electric field strength $E^+$. In the appendix of~\cite{Li07}, it is shown that precisely the same degree of ionization $\sigma^-=\sigma^-(E^+)$ from Eq.~(\ref{edensminfty}) is reached behind a planar front even if it does not propagate uniformly, if only the electric field $E^+$ is the same at the moment when the electron discontinuity passes the point of observation. Behind a slightly curved front, the degree of ionization will be approximately the same.

As a consequence, the conductivity further behind the front varies as a function of evolution history, as the immobile ions with their positive charge act as a memory. E.g., when a streamer emerges from an avalanche in a homogeneous background field ${\bf E}_{\text{back}}$, the conductivity in the evolving channel will increase from $\sigma^-(|{\bf E}_{\text{back}}|)$ immediately after the avalanche phase to $\sigma^-(|{\bf E}_{\text{enh}}(t)|)$ where ${\bf E}_{\text{enh}}(t)$ is the enhanced field at the tip of the streamer at time $t$. This increasing electron and ion density along the axis of the streamer finger is typically seen in simulations, e.g., in Fig.~1 of~\cite{CaroJPD} for an emerging streamer or in Fig.~3(a) of~\cite{CaroPRE} for a fully developed long streamer.

When an electric field is applied to a body with spatially varying conductivity, in general, space charges can form in the interior, since charge conservation $\partial_t q+\nabla\cdot{\bf j}=0$ with $q=\rho-\sigma$ and ${\bf j}=\sigma{\bf E}$ can be rewritten in the form
\begin{equation}
\label{Maxwell} \left(\partial_t+\sigma\right)\, (\rho-\sigma)+{\bf E}\cdot\nabla \sigma=0,
\end{equation}
in agreement with (\ref{PDE1})--(\ref{PDE3}), i.e., the space charge $\rho-\sigma$ will decay on the Maxwell time scale $1/\sigma$ only if there is no conductivity gradient $\nabla\sigma$ in the direction parallel to the electric field ${\bf E}$. However, in practice such space charge effects are typically small if the evolution is slow and consequently $|\nabla\sigma|$ is small.

\subsection{The approximation of ideal conductivity}

Lozansky and Firsov in their Russian textbook and in their short English article~\cite{loza73} suggested to neglect the weak field in the ionized interior completely together with possible weak space charge effects, and to approximate the streamer interior as ideally conducting, i.e., to assume infinite conductivity and
\begin{equation}
\phi={\rm const} \quad \text{in} \quad \Omega^-.
\end{equation}
Inspection of simulations show that this is a valid approximation for the weak interior fields in comparison to the strong fields in the exterior. Therefore the approximation was adapted in the
model~\cite{PRL05,SIAM06,brau07,saleh08}, and its validity was further tested on simulations in section IV of~\cite{brau07}.

\subsection{The approximation of constant conductivity}

A shortcoming of the ideal conductivity approximation is that it does not allow for a potential drop along the length of the streamer which is an important observable in experiments. In fact, the electric field along the axis of a long streamer is frequently assumed to be constant; it is an important point of scientific discussion and an ingredient of coarse grained models for complete streamer trees.

The approximation of constant conductivity in the interior is justified if the width of the streamer and the field enhancement at its head vary relatively slowly during propagation as it is the case in many simulations. We therefore here adapt the approximation of constant conductivity $\sigma^-$ in the ionized streamer interior. According to Eq.~(\ref{Maxwell}), space charges $\rho-\sigma$ then decay on the time scale $1/\sigma^-$. For a front propagating with velocity $v$, this corresponds to a spatial decay on the length $\ell_\alpha^-=v/\sigma^-$ (\ref{lambdaminus1}), generalizing the explicit asymptotic behavior (\ref{asympplanar}) derived for uniformly translating planar fronts. Behind this layer, there are no space charges and the Laplace equation approximates the ionized interior as well:
\begin{equation}
\label{Laplace-} \nabla^2\phi=0, \quad {\bf E}=-\nabla\phi \quad \text{in} \quad \Omega^-.
\end{equation}

From a mathematical point of view, a finite conductivity in the interior might lead to an additional
regularization mechanism. We have a model with two Laplacian fields on both side of the interface. This type of model is called Muskat, or Muskat-Leibenzon, problem in hydrodynamics and describes the evolution of the interface between two immiscible fluids in a porous medium or Hele-Shaw cell under applied pressure gradients or fluid injection/extraction. The so-called Hele-Shaw problem is thus the one-phase version of the Muskat problem, see for example \cite{musk34,jian90,howi00,sieg04}.

\section{Conclusion and outlook}
\label{conclusion}

We have derived the curvature corrections to a moving boundary approximation of streamer ionization fronts within the minimal model. We also have introduced a constant conductivity approximation in the ionized streamer interior. The matching procedure between inner and outer region followed the standard procedure at the back end of the front. However, ahead of the front, the matching region shrunk to a line. The solvability analysis was based on integrating from $-\infty$ up to this line. It involved three zero modes of the adjoint operator, one of them diverging. Nevertheless, the boundary approximation could be carried through, and even fully analytically.

The analysis has two important merits. First, the curvature correction to the potential jump $\phi^+-\phi^-$ across the interface might remove non-regular solutions \cite{Prok08} found in the ``minimal regularized boundary model" \cite{PRL05,SIAM06,brau07,saleh08}. Second, it provides a very practical relation between the front curvature and the electric field behind the front.

The derived boundary approximation accounts for propagating streamers in fairly homogeneous fields and far from electrodes. On the other hand, ionization fronts driven by monopole fields (e.g., close to sharp needle electrodes) or during branching undergo fast temporal changes of the electric field $E^+$ on the ionization front. They cannot be approximated by uniformly translating fronts anymore; and time dependent corrections need to be taken into account. In a forthcoming paper, we will  explore how far charge conservation principles will carry us towards a full derivation of a moving boundary approximation for the minimal streamer model.

\begin{acknowledgments}
We acknowledge inspiring discussions with W.~van Saarloos. The work of F.B. was
supported by The Netherlands Organisation for Scientific Research (NWO) through Contract No. 633.000.401 within the program ``Dynamics of Patterns,'' and the guest visit of B.D. to CWI Amsterdam by NDNS+ cluster.
\end{acknowledgments}

\appendix

\section{Solving the zero modes equations}
\label{appendix2}

In this appendix, we construct the exact expression for the zero modes. The
system of ODE we need to solve for the zero modes is, according to
Eqs.~(\ref{eq0mode1b})-(\ref{eq0mode4b}) the following
\begin{eqnarray}
\label{eq0mode1bb}
(f_0-\sigma_0)\psi_{\sigma}+f_0\psi_{\rho} -\psi_E -(v_0+E_0)\partial_{\zeta}\psi_{\sigma}&=&0, \\
\label{eq0mode2bb}
\sigma_0\psi_{\sigma}+\psi_E-v_0\partial_{\zeta}\psi_{\rho}&=&0, \\
\label{eq0mode3bb}
\sigma_0 f_0' (\psi_{\sigma}+\psi_{\rho})-(\partial_{\zeta}\sigma_0)\psi_{\sigma} + c_{\phi} &=&\partial_{\zeta}\psi_E.
\end{eqnarray}

Summing Eq.~(\ref{eq0mode1bb}) with Eq.~(\ref{eq0mode2bb}) leads to
\begin{equation}
    \label{psisigmasol}
E_0\partial_{\zeta}\psi_{\sigma}=-v_0\partial_{\zeta}(\psi_{\sigma}+\psi_{\rho})+f_0(\psi_{\sigma}+\psi_{\rho}).
\end{equation}
We will derive below an equation for $\psi_{\sigma}+\psi_{\rho}$. Once we get
it, the complete system is solved with the help of (\ref{psisigmasol}) and
(\ref{eq0mode3bb}). For this purpose,
we take the derivative of Eq.~(\ref{eq0mode2bb}) and replace
$\partial_{\zeta}\psi_E$ by its expression given by (\ref{eq0mode3bb}) to get
\begin{equation}
    \label{tmp1}
    \sigma_0\partial_{\zeta}\psi_{\sigma}+\sigma_0 f_0'(\psi_{\sigma}+\psi_{\rho})+c_{\phi}- v_0\partial_{\zeta}^2\psi_{\rho}=0
\end{equation}
Taking the derivative of (\ref{psisigmasol}) and using the fact that (see
Eqs.~(\ref{chargeplanar}) and (\ref{deff0}))
\begin{equation}
    \partial_{\zeta} f_0=-f_0' \partial_{\zeta}E_0 \quad \text{and} \quad \partial_{\zeta}E_0=\frac{E_0\, \sigma_0}{v_0},
\end{equation}
we obtain
\begin{eqnarray}
    \label{tmp2}
&-&v_0\partial_{\zeta}^2(\psi_{\sigma}+\psi_{\rho})-\frac{E_0\sigma_0 f_0'}{v_0} (\psi_{\sigma}+\psi_{\rho})+f_0\partial_{\zeta}(\psi_{\sigma}+\psi_{\rho}) \nonumber \\ &-&\frac{E_0\sigma_0}{v_0}\partial_{\zeta}\psi_{\sigma}-E_0\partial_{\zeta}^2\psi_{\sigma}=0.
\end{eqnarray}
We finally multiply (\ref{tmp1}) by $E_0/v_0$ and sum it with (\ref{tmp2}) to
get the equation for $\psi_{\sigma}+\psi_{\rho}$
\begin{equation}
    \label{psis+psirsol}
-
(v_0+E_0)\partial_{\zeta}^2(\psi_{\sigma}+\psi_{\rho})+f_0\partial_{\zeta}(\psi_{\sigma}+\psi_{\rho})+
    \frac{c_{\phi}E_0}{v_0}=0.
\end{equation}
This is formally a first order ODE and can be solved exactly. Once
$\psi_{\sigma}+\psi_{\rho}$ is known, $\psi_{\sigma}$ is computed with
(\ref{psisigmasol}) and then $\psi_E$ is obtained from (\ref{eq0mode3bb}).

\subsection{Expression of $\psi_{\sigma}+\psi_{\rho}$}
\label{appendix2b}

To solve (\ref{psis+psirsol}), we use the ansatz
\begin{eqnarray}
    \partial_{\zeta}(\psi_{\sigma}+\psi_{\rho})(\zeta)&=&C(\zeta) e^{g(\zeta)},\quad \text{with} \\
    \label{eqforg}
    g(\zeta)&=&\int_{-\infty}^{\zeta} dx \frac{f_0(x)}{v_0+E_0(x)}.
\end{eqnarray}
The equation for $C$ reads
\begin{equation}
    \partial_{\zeta}C=\frac{c_{\phi}E_0}{v_0(v_0+E_0)}e^{-g(\zeta)}.
\end{equation}
After integration we obtain,
\begin{equation}
	\label{cczeta}
    C(\zeta)=C(-\infty)+\int_{-\infty}^{\zeta}dx\frac{c_{\phi}E_0}{v_0(v_0+E_0)}e^{-g(x)}.
\end{equation}
However from the asymptotic analysis of the system of ODE for the zero mode (see (\ref{asymppsisdir}) and (\ref{asymppsirdir})), we know that
\begin{equation}
    C(-\infty)=\partial_{\zeta}(\psi_{\sigma}+\psi_{\rho})(\zeta_{\text{ini}}\to -\infty)=0.
\end{equation}
The solution we search takes thus the form
\begin{equation}
    \label{psis+psirtemp}
    (\psi_{\sigma}+\psi_{\rho})(\zeta)=c_{\sigma \rho}+\int_{-\infty}^{\zeta}dy\, e^{g(y)} \int_{-\infty}^{y}dx
\frac{c_{\phi}E_0}{v_0(v_0+E_0)}e^{-g(x)},
\end{equation}
where
\begin{equation}
    c_{\sigma \rho}=(\psi_{\sigma}+\psi_{\rho})(-\infty).
\end{equation}

The explicit form of $g(\zeta)$ is computed using (\ref{eqforg})
\begin{eqnarray}
    g(\zeta)&=&\int_{-\infty}^{\zeta} dx \frac{f_0(x)}{v_0+E_0(x)}, \nonumber
\\
            &=&-\int_{-\infty}^{\zeta} dx \frac{E_0(x)
\alpha_0(x)}{v_0+E_0(x)}.
\end{eqnarray}
Using the change of variable
\begin{eqnarray}
    \label{changevariable}
    y&=&E_0(x) \\
    dy&=&\partial_{x}E_0(x) \, dx=\frac{E_0(x) \rho_0(x)}{v_0+E_0(x)} dy,
\end{eqnarray}
we obtain
\begin{eqnarray}
    g(\zeta)&=&-\int_{0}^{E_0(\zeta)}dx
\frac{\alpha(|x|)}{\rho(x)}=\int_{0}^{|E_0(\zeta)|}dx \frac{\alpha(x)}{\rho(x)},
\nonumber \\
            &=&-\int_{0}^{|E_0(\zeta)|}dx
\frac{\rho'(x)}{\rho(x)}=\ln\left(\frac{\rho(0)}{\rho(|E_0(\zeta)|)} \right),
\nonumber \\
\label{eqforg2}
            &=&\ln\left(\frac{\sigma^-}{\rho_0(\zeta)} \right),
\end{eqnarray}
where we use the notation $\rho(|E_0(\zeta)|)= \rho(E_0(\zeta))\equiv \rho_0(\zeta)$ in this Appendix. Substituting Eq.~(\ref{eqforg2}) in Eq.~(\ref{psis+psirtemp}) we obtain
\begin{equation}
    (\psi_{\sigma}+\psi_{\rho})(\zeta)=c_{\sigma \rho}+\int_{-\infty}^{\zeta}
\frac{dy}{\rho_0(y)} \int_{-\infty}^{y}dx
\frac{c_{\phi}E_0}{v_0(v_0+E_0)}\rho_0(x),
\end{equation}
With the change of variable (\ref{changevariable}) we obtain
\begin{equation}
    (\psi_{\sigma}+\psi_{\rho})(\zeta)=c_{\sigma
\rho}+\frac{c_{\phi}}{v_0}\int_{0}^{E_0(\zeta)}dx\, \frac{s\left(E_0^{-
1}(x)\right)(v_0+x)}{x\rho(x)^2},
\end{equation}
where
\begin{equation}
    s\left(E_0^{-1}(x)\right)=\int_{-\infty}^{E_0^{-1}(x)}dy
\frac{E_0(y)\rho_0(y)}{v_0+E_0(y)}
\end{equation}
Again, with the change of variable (\ref{changevariable}) one can compute that
\begin{equation}
    s\left(E_0^{-1}(x)\right)=\int_{0}^{x}dz = x
\end{equation}
Consequently, the final form for the solution of Eq.~(\ref{psis+psirsol}) is
given by
\begin{equation}
    \label{psis+psir}
(\psi_{\sigma}+\psi_{\rho})(\zeta)=c_{\sigma
\rho}+\frac{c_{\phi}}{v_0}\int_{0}^{E_0(\zeta)}dx\, \frac{(v_0+x)}{\rho(x)^2},
\end{equation}

\subsection{Expression of $\psi_{\sigma}$}
\label{appendix2c}

We have now to solve Eq.~(\ref{psisigmasol}). We rewrite it using the exact
solution (\ref{psis+psir}). In particular we have
\begin{eqnarray}
\partial_{\zeta}(\psi_{\sigma}+\psi_{\rho})&=&\frac{c_{\phi}}{v_0}\frac{v_0+E_0}
{\rho_0^2}\partial_{\zeta}E_0 \\
    &=&\frac{c_{\phi}}{v_0} \frac{E_0}{\rho_0}
\end{eqnarray}
Thus
\begin{equation}
    \label{derpsis}
    \partial_{\zeta}\psi_{\sigma}=-\frac{c_{\phi}}{\rho_0}-
\alpha_0(\psi_{\sigma}+\psi_{\rho})
\end{equation}
Remark that it is easy to verify with the help of this relation that we can
recover the asymptotic behavior (\ref{asymppsisdir}). After an integration of
(\ref{derpsis}) we obtain
\begin{eqnarray}
\psi_{\sigma}&=&\psi_{\sigma}(\zeta_{\text{ini}})-
c_{\phi}\int_{\zeta_{\text{ini}}}^{\zeta}\frac{dx}{\rho_0(x)} \nonumber \\ &-
&\int_{\zeta_{\text{ini}}}^{\zeta}dx\,  \alpha_0(x)\left[c_{\sigma
\rho}+\frac{c_{\phi}}{v_0}\int_{0}^{E_0(x)}dy\, \frac{(v_0+y)}{\rho(y)^2}\right]
\end{eqnarray}
From the asymptotic analysis of (\ref{derpsis}), and as we show explicitly
below, one can verify that only the first integral of this last expression
diverges when $\zeta_{\text{ini}} \to -\infty$. Using the change of variable
(\ref{changevariable}) we finally get
\begin{eqnarray}
    \label{psis}
\psi_{\sigma}(\zeta)&=&\psi_{\sigma}(\zeta_{\text{ini}})-
c_{\phi}\int_{|E_0(\zeta_{\text{ini}})|}^{|E_0(\zeta)|}dx \frac{v_0-
x}{x\rho(x)^2} \nonumber \\ &-&c_{\sigma
\rho}\int_{|E_0(\zeta_{\text{ini}})|}^{|E_0(\zeta)|}dx \frac{(v_0-
x)\alpha(x)}{x\rho(x)} \nonumber \\
&+&\frac{c_{\phi}}{v_0}\int_{|E_0(\zeta_{\text{ini}})|}^{|E_0(\zeta)|}dx
\frac{(v_0-x)\alpha(x)}{x\rho(x)}\int_0^x dy\frac{v_0-y}{\rho(y)^2}.
\end{eqnarray}
In this last expression only the first integral diverges when
$\zeta_{\text{ini}} \to -\infty$. The singularity can be explicitly extract
thanks to the relation
\begin{eqnarray}
\label{psis-zetain}
&&\int_{|E_0(\zeta_{\text{ini}})|}^{|E_0(\zeta)|}dx \frac{v_0-
x}{x\rho(x)^2}=\int_{|E_0(\zeta_{\text{ini}})|}^{|E_0(\zeta)|}dx \frac{v_0-
x}{x\rho(x)^2}+\frac{\zeta_{\text{ini}}-\zeta}{\sigma^-}\nonumber \\  &+&
\frac{\zeta-\zeta_{\text{ini}}}{\sigma^-} \nonumber \\
&=&\int_{|E_0(\zeta_{\text{ini}})|}^{|E_0(\zeta)|}dx \frac{v_0-x}{x\rho(x)^2}-
\frac{1}{\sigma^-}\int_{|E_0(\zeta_{\text{ini}})|}^{|E_0(\zeta)|}dx \frac{v_0-
x}{x\rho(x)} 
\nonumber \\ &+& \frac{\zeta-\zeta_{\text{ini}}}{\sigma^-} \nonumber \\
&=&\int_{|E_0(\zeta_{\text{ini}})|}^{|E_0(\zeta)|}dx\left[\frac{1}{\rho(x)}-
\frac{1}{\sigma^-} \right]\frac{v_0-x}{x\rho(x)}+\frac{\zeta-
\zeta_{\text{ini}}}{\sigma^-},
\end{eqnarray}
where we used Eq.~(\ref{E}). The integral of this last relation is always
finite. Expression (\ref{psis}) can then be written as in the main text
\begin{equation}
\label{psisshortapp}
\psi_{\sigma}(\zeta) = \frac{c_{\phi}}{\sigma^-} \zeta_{\text{ini}} +
a_{\sigma}(\zeta) + b_{\sigma}(\zeta) e^{\lambda \zeta_{\text{ini}}}\ ,
\end{equation}
where
\begin{eqnarray}
    \label{express-asig}
    a_{\sigma}(\zeta)&=&\psi_{\sigma}(\zeta_{\text{ini}})-
c_{\phi}\int_{0}^{|E_0(\zeta)|}dx\left[\frac{1}{\rho(x)}-\frac{1}{\sigma^-}
\right]\frac{v_0-x}{x\rho(x)} \nonumber \\ &-&\frac{c_{\phi}}{\sigma^-}\zeta -
c_{\sigma \rho}\int_{0}^{|E_0(\zeta)|}dx \frac{(v_0-x)\alpha(x)}{x\rho(x)}
\nonumber \\ &+&\frac{c_{\phi}}{v_0}\int_{0}^{|E_0(\zeta)|}dx \frac{(v_0-
x)\alpha(x)}{x\rho(x)}\int_0^x dy\frac{v_0-y}{\rho(y)^2}.
\end{eqnarray}
The expression $b_{\sigma}(\zeta)$ is not useful for our analyze since the
contribution of this term in Eq.~(\ref{psisshortapp}) can be made as small as we
want since $|\zeta_{\text{ini}}|$ can be as large as we want (but finite).

\subsection{Expression of $\psi_E$}
\label{appendix2d}

Equation (\ref{eq0mode3bb}) can be written as
\begin{eqnarray}
    \partial_{\zeta}\psi_E&=&\sigma_0 f_0'(\psi_{\sigma}+\psi_{\rho})-
(\partial_{\zeta}\sigma_0) \psi_{\sigma}+c_{\phi} \nonumber \\
        &=&\sigma_0
f_0'(\psi_{\sigma}+\psi_{\rho})+\sigma_0\partial_{\zeta}\psi_{\sigma}-
\partial_{\zeta}(\sigma_0\psi_{\sigma})+c_{\phi} \nonumber \\
        &=&\sigma_0 f_0'(\psi_{\sigma}+\psi_{\rho})-
\sigma_0\frac{c_{\phi}}{\rho_0}-\sigma_0\alpha_0(\psi_{\sigma}+\psi_{\rho})
\nonumber \\ &-&\partial_{\zeta}(\sigma_0\psi_{\sigma})+c_{\phi}
\end{eqnarray}
where we used the expression (\ref{derpsis}) for
$\partial_{\zeta}\psi_{\sigma}$. Using the relation (\ref{deff0p}) for $f_0'$,
this last expression reduces to
\begin{equation}
\label{temp3}   \partial_{\zeta}\psi_E=-
\frac{\sigma_0\alpha_0}{E_0}(\psi_{\sigma}+\psi_{\rho})-
\frac{c_{\phi}v_0}{v_0+E_0}-\partial_{\zeta}(\sigma_0\psi_{\sigma})+c_{\phi},
\end{equation}
where we used also the relation (\ref{sigma}) between $\sigma_0$ and $\rho_0$.
It is also easy to see that in the limit $\zeta \to -\infty$ we recover the
asymptotic form (\ref{asymppsiedir}). The expression of $\psi_E$ is obtained
after an integration of (\ref{temp3})
\begin{eqnarray}
    \psi_E(\zeta)&=&\psi_E(\zeta_{\text{ini}})-
\int_{\zeta_{\text{ini}}}^{\zeta}dx \frac{\sigma_0\alpha_0}{E_0}\left[c_{\sigma
\rho}\right. \nonumber \\  &+& \left. \frac{c_{\phi}}{v_0}\int_{0}^{E_0(x)}dy\,
\frac{(v_0+y)}{\rho(y)^2}\right]  -\sigma_0(\zeta)\psi_{\sigma}(\zeta)\nonumber
\\ &+&\sigma_0(\zeta_{\text{ini}})\psi_{\sigma}(\zeta_{\text{ini}})-
c_{\phi}v_0\int_{\zeta_{\text{ini}}}^{\zeta}\frac{dx}{v_0+E_0} \nonumber \\
&+&c_{\phi}(\zeta-\zeta_{\text{ini}})
\end{eqnarray}
Using the change of variable (\ref{changevariable}) we get
\begin{eqnarray}
\label{temp4}
    \psi_E(\zeta)&=&\psi_E(\zeta_{\text{ini}})-
v_0\int_{E_0(\zeta_{\text{ini}})}^{E_0(\zeta)}dx
\frac{\alpha(|x|)}{x^2}\left[c_{\sigma \rho}\right. \nonumber \\  &+& \left.
\frac{c_{\phi}}{v_0}\int_{0}^{x}dy\, \frac{(v_0+y)}{\rho(y)^2}\right]  -
\sigma_0(\zeta)\psi_{\sigma}(\zeta)\nonumber \\
&+&\sigma_0(\zeta_{\text{ini}})\psi_{\sigma}(\zeta_{\text{ini}})-
c_{\phi}v_0\int_{E_0(\zeta_{\text{ini}})}^{E_0(\zeta)}\frac{dx}{x\rho(x)}
\nonumber \\ &+&c_{\phi}(\zeta-\zeta_{\text{ini}})
\end{eqnarray}
The two last terms diverge when $\zeta_{\text{ini}} \to -\infty$ but their sum
is finite. With the help of the definition (\ref{E}) we can write this sum as a
finite quantity. Indeed, we know that
\begin{equation}
    \zeta-\zeta_{\text{ini}}=\int_{E_0(\zeta_{\text{ini}})}^{E_0(\zeta)}dx
\frac{v_0+x}{x\rho(x)}
\end{equation}
and we substitute it in (\ref{temp4}) to get
\begin{eqnarray}
&&  \psi_E(\zeta)=\psi_E(\zeta_{\text{ini}})+c_{\sigma
\rho}v_0\int_{|E_0(\zeta_{\text{ini}})|}^{|E_0(\zeta)|}\frac{\alpha(x)}{x^2}
\nonumber \\ &-&c_{\phi}\int_{|E_0(\zeta_{\text{ini}})|}^{|E_0(\zeta)|}dx
\frac{\alpha(x)}{x^2}\int_{0}^{x}dy \frac{(v_0+y)}{\rho(y)^2} -
\sigma_0(\zeta)\psi_{\sigma}(\zeta)\nonumber \\
&+&\sigma_0(\zeta_{\text{ini}})\psi_{\sigma}(\zeta_{\text{ini}})-
c_{\phi}\int_{|E_0(\zeta_{\text{ini}})|}^{|E_0(\zeta)|}\frac{dx}{\rho(x)}.
\end{eqnarray}
The only part of this relation that diverge when $\zeta_{\text{ini}} \to -
\infty$ is contained in the expression of $\psi_{\sigma}(\zeta)$. We can still
simplify this last relation by using the following formulas
\begin{eqnarray}
\int_{|E_0(\zeta_{\text{ini}})|}^{|E_0(\zeta)|}\frac{\alpha(x)}{x^2}&=&[\alpha(x
)]_{|E_0(\zeta_{\text{ini}})|}^{|E_0(\zeta)|} \nonumber \\
&=&\alpha(|E_0(\zeta)|)-\alpha(|E_0(\zeta_{\text{ini}})|).
\end{eqnarray}
An important point to emphasize is that this calculation is the only one so far
where the explicit form of
the function $\alpha(x)=e^{-1/x}$ was used to simplify the expression. The
second relation is obtained after an integration by parts
\begin{eqnarray}
    &&\int_{|E_0(\zeta_{\text{ini}})|}^{|E_0(\zeta)|}dx
\frac{\alpha(x)}{x^2}\int_{0}^{x}dy \frac{(v_0+y)}{\rho(y)^2}= \nonumber \\
&&(\alpha(|E_0(\zeta)|)-
\alpha(|E_0(\zeta_{\text{ini}})|))\int_{|E_0(\zeta_{\text{ini}})|}^{|E_0(\zeta)|
}dx \frac{v_0-x}{\rho(x)^2}\nonumber \\&-
&\int_{|E_0(\zeta_{\text{ini}})|}^{|E_0(\zeta)|}dx \frac{\alpha(x)(v_0-
x)}{\rho(x)^2}.
\end{eqnarray}

Finally we get
\begin{eqnarray}
    \label{psie}
\psi_E(\zeta)&=&c_{\zeta_{\text{ini}}}+c_{\sigma \rho}v_0(\alpha_0(\zeta)-
\alpha_0(\zeta_{\text{ini}})) \nonumber \\ &-&c_{\phi}(\alpha_0(\zeta)-
\alpha_0(\zeta_{\text{ini}}))\int_{|E_0(\zeta_{\text{ini}})|}^{|E_0(\zeta)|}dx
\frac{v_0-x}{\rho(x)^2} \nonumber \\ &+&c_{\phi}
\int_{|E_0(\zeta_{\text{ini}})|}^{|E_0(\zeta)|}dx \frac{\alpha(x)(v_0-
x)}{\rho(x)^2}-\sigma_0(\zeta)\psi_{\sigma}(\zeta)\nonumber \\ &-
&c_{\phi}\int_{|E_0(\zeta_{\text{ini}})|}^{|E_0(\zeta)|}\frac{dx}{\rho(x)},
\end{eqnarray}
with
\begin{equation}
    \label{czeta}
    c_{\zeta_{\text{ini}}}=\psi_E(\zeta_{\text{ini}})+\sigma_0(\zeta_{\text{in
i}})\psi_{\sigma}(\zeta_{\text{ini}})
\end{equation}
Expression (\ref{psie}) can then be written as in the main text
\begin{equation}
\label{psieshortapp}
\psi_{E}(\zeta) = -\frac{c_{\phi}}{\sigma^-} \zeta_{\text{ini}} +
a_{E}(\zeta) + b_{E}(\zeta) e^{\lambda \zeta_{\text{ini}}}\ ,
\end{equation}
where
\begin{eqnarray}
    \label{express-aE}
    a_{E}(\zeta)&=&\tilde{c}_{\zeta_{\text{ini}}}+c_{\sigma \rho}v_0
\alpha_0(\zeta) -c_{\phi}\alpha_0(\zeta)\int_{0}^{|E_0(\zeta)|}dx \frac{v_0-
x}{\rho(x)^2} \nonumber \\ &+&c_{\phi} \int_{0}^{|E_0(\zeta)|}dx
\frac{\alpha(x)(v_0-x)}{\rho(x)^2}-
\sigma_0(\zeta)a^{(i)}_{\sigma}(\zeta)\nonumber \\ &-
&c_{\phi}\int_{0}^{|E_0(\zeta)|}\frac{dx}{\rho(x)},
\end{eqnarray}
where we used (\ref{psisshortapp}) and where
\begin{equation}
    \label{czetatilde}
    \tilde{c}_{\zeta_{\text{ini}}}=\psi_E(\zeta_{\text{ini}})+\sigma^-
\psi_{\sigma}(\zeta_{\text{ini}}).
\end{equation}
As for (\ref{psisshortapp}), the expression $b_{E}(\zeta)$ is not useful for our
analyze since the contribution of this term in Eq.~(\ref{psieshortapp}) can be
made as small as we want.

\subsection{Expression of $A$}
\label{app:expressA}

Since we know that $E_0$, $\sigma_0$ and $\rho_0$ decay exponentially for $\zeta
\to -\infty$, see (\ref{asympplanar}), and since the zero modes diverge at most
linearly in this limit, we can already replace $\zeta_{\text{c}}$ by $-\infty$
to compute $A$ and $B$.

We know the exact expression of the zero modes and we start to compute the
relevant quantities for the relation (\ref{BC2}) which will give us the boundary
conditions we search for. We use in this section the notation
\begin{eqnarray}
    \label{xtemp1}
    (\psi_{\sigma}+\psi_{\rho})(\zeta)&=&c_{\sigma \rho}+c_{\phi}
h(E_0(\zeta)), \\
    h(E_0(\zeta))&=&\frac{1}{v_0}\int_{0}^{E_0(\zeta)}dx\,
\frac{(v_0+x)}{\rho(x)^2}.
\end{eqnarray}
From the definition (\ref{eqA}) of $A$ and Eq.~(\ref{xtemp1}), we have
\begin{eqnarray}
    \label{tempa1}
    A&=&\int_{-\infty}^0 d\zeta \left[ \psi_{\sigma} \partial_{\zeta}\sigma_0
+\left(c_{\sigma \rho}+c_{\phi} h-\psi_{\sigma} \right)\partial_{\zeta}\rho_0
\right] \nonumber \\
    &=&-\sigma^- c_{\sigma \rho}+ \int_{-\infty}^0 d\zeta \,
\psi_{\sigma}\left(\partial_{\zeta}\sigma_0-\partial_{\zeta}\rho_0 \right)
\nonumber \\ &+&c_{\phi}\int_{-\infty}^0 d\zeta\, h\, \partial_{\zeta}\rho_0
\end{eqnarray}
Using (\ref{chargeplanar}) we can compute the derivation of the ion density of
the planar front solution
\begin{eqnarray}
    \partial_{\zeta}\rho_0&=&\partial_{\zeta}\left(\frac{v_0+E_0}{v_0}\sigma_0
\right), \nonumber \\
&=&\frac{E_0\sigma_0^2}{v_0^2}+\frac{v_0+E_0}{v_0}\partial_{\zeta}\sigma_0.
\end{eqnarray}
We substitute it in Eq.~(\ref{tempa1}) to obtain
\begin{eqnarray}
    \label{tempa2}
    A&=&-\sigma^- c_{\sigma \rho}- \int_{-\infty}^0 d\zeta \, \psi_{\sigma}
\frac{E_0\sigma_0^2}{v_0^2}- \int_{-\infty}^0 d\zeta \, \frac{\psi_{\sigma} E_0
\partial_{\zeta}\sigma_0}{v_0}\nonumber \\ &+&c_{\phi}\int_{-\infty}^0 d\zeta\,
h\, \partial_{\zeta}\rho_0
\end{eqnarray}
The two last integral can be simplified. Indeed, using an integration by parts
we have
\begin{eqnarray}
    \label{tempa3}
    &&\int_{-\infty}^0 d\zeta \, \psi_{\sigma} E_0
\partial_{\zeta}\sigma_0=\left[ \psi_{\sigma} E_0 \sigma_0\right]_{-\infty}^0
\nonumber \\ &-&\int_{-\infty}^0 d\zeta \,\sigma_0\left( E_0
\partial_{\zeta}\psi_{\sigma}+ \psi_{\sigma}\partial_{\zeta}E_0\right),
\end{eqnarray}
where we have still to replace $\partial_{\zeta}\psi_{\sigma}$ and
$\partial_{\zeta}E_0$ by their expression (see (\ref{derpsis}) and
(\ref{chargeplanar})). The last integral of (\ref{tempa2}) can also be
simplified by an integration by parts
\begin{eqnarray}
    \label{tempa4}
    \int_{-\infty}^0 d\zeta\, h\, \partial_{\zeta}\rho_0=- \int_{-\infty}^0
d\zeta\, \frac{E_0}{v_0}
\end{eqnarray}
Substituting Eqs.~(\ref{tempa3}) and (\ref{tempa4}) in Eq.~(\ref{tempa2}),
making the change of variable (\ref{changevariable}) and using the definition
(\ref{edensminfty}) of $\sigma^-$, we get
\begin{eqnarray}
    A&=&\sigma_0(0)\psi_{\sigma}(0)+c_{\phi}\int_{0}^{E^+}\frac{dx}{\rho(x)}
\\ &+&c_{\phi}\int_{0}^{E^+}dx \, \alpha(x) h(-
x)+\frac{c_{\phi}}{v_0}\int_{0}^{E^+}dx\frac{v_0-x}{\rho(x)}, \nonumber
\end{eqnarray}
Notice that the only term that diverge when $\zeta_{\text{ini}} \to -\infty$ is
contained into $\psi_{\sigma}(0)$. Now this formula can be simplified further
since actually the two last integrals cancel. Indeed, with an integration by
parts we get
\begin{eqnarray}
    &&c_{\phi}\int_{0}^{E^+}dx \, \alpha(x)h(-x)=-
\frac{c_{\phi}}{v_0}\left[H(x) \int_0^x dy\, \frac{v_0-
y}{\rho(y)^2}\right]_0^{E^+} \nonumber \\
    &+&\frac{c_{\phi}}{v_0}\int_0^{E^+}dx \frac{v_0-x}{\rho(x)^2}(-
\rho(x)+H(E^+)),
\end{eqnarray}
where we used the notation
\begin{eqnarray}
    H(x)=\int dx\, \alpha(x)&=&-\int_x^{E^+} dy \, \alpha(y)+H(E^+), \nonumber
\\
                          &=&-\rho(x)+H(E^+).
\end{eqnarray}
It is easy to verify that $H(0)=0$, consequently we have
\begin{equation}
    c_{\phi}\int_{0}^{E^+}dx\, \alpha(x)h(-x)=-
\frac{c_{\phi}}{v_0}\int_0^{E^+}dx \frac{v_0-x}{\rho(x)}.
\end{equation}
The final expression of $A$ is then rather simple and reads
\begin{equation}
    \label{expressionA}
    A=\sigma_0(0)\psi_{\sigma}(0)+c_{\phi}\int_{0}^{E^+}\frac{dx}{\rho(x)}.
\end{equation}
This expression can be written as in the main text
\begin{equation}
    A=\frac{\sigma_0(0)c_{\phi}}{\sigma^-} \zeta_{\text{ini}}+ a_A+b_A
e^{\lambda \zeta_{\text{ini}}},
\end{equation}
where
\begin{equation}
    \label{express-aA}
    a_A=\sigma_0(0)a_{\sigma}(0)+c_{\phi}\int_{0}^{E^+}\frac{dx}{\rho(x)}
\end{equation}

\subsection{Expression of $A+\psi_E(0)$}
\label{app:expressA+psie}

We just need to sum the expression (\ref{expressionA}) of $A$ with the
expression (\ref{psie}) of $\psi_E$ evaluated at $\zeta=0$. We get
\begin{eqnarray}
\label{A+psie0}
&&A+\psi_E(0)=c_{\zeta_{\text{ini}}} +c_{\sigma \rho}v_0(\alpha(E^+)-
\alpha_0(\zeta_{\text{ini}})) \nonumber \\ &-&c_{\phi}(\alpha(E^+)-
\alpha_0(\zeta_{\text{ini}}))\int_{|E_0(\zeta_{\text{ini}})|}^{E^+}dx \frac{v_0-
x}{\rho(x)^2} \nonumber \\ &+&c_{\phi} \int_{|E_0(\zeta_{\text{ini}})|}^{E^+}dx
\frac{\alpha(x)(v_0-x)}{\rho(x)^2},
\end{eqnarray}
with $c_{\zeta_{\text{ini}}}$ defined by (\ref{czeta}). This expression can be
written as in the main text
\begin{equation}
    A+\psi_E(0)=a_A+a_E(0)+b_E(0) e^{\lambda \zeta_{\text{ini}}},
\end{equation}
where
\begin{eqnarray}
\label{aA+aE}
&&a_A+a_E(0)=\tilde{c}_{\zeta_{\text{ini}}} +c_{\sigma \rho}v_0\alpha(E^+)
\nonumber \\ &-&c_{\phi}\alpha(E^+)\int_{0}^{E^+}dx \frac{v_0-x}{\rho(x)^2}
\nonumber \\ &+&c_{\phi} \int_{0}^{E^+}dx \frac{\alpha(x)(v_0-x)}{\rho(x)^2},
\end{eqnarray}
and where $\tilde{c}_{\zeta_{\text{ini}}}$ is defined by (\ref{czetatilde}).

\subsection{Expression of $B$}
\label{app:expressB}

Using the change of variables (\ref{changevariable}) in the definition
(\ref{eqB}) of $B$ we get
\begin{equation}
    B=-\int_0^{E^+} dx \frac{v_0-x}{\rho(x)}\psi_E\left(E_0^{-1}(-x) \right).
\end{equation}
Replacing $\psi_E$ by its exact form (\ref{psie}) we get
\begin{eqnarray}
    \label{btemp}
    B&=&-\int_0^{E^+} dx \frac{v_0-
x}{\rho(x)}\left[c_{\zeta_{\text{ini}}}+c_{\sigma \rho}v_0(\alpha(x)-
\alpha(|\zeta_{\text{ini}})|) \right. \nonumber \\ &-& \frac{v_0}{v_0-x}\rho(x)
\psi_{\sigma}\left(E_0^{-1}(-x)\right) \nonumber \\ &-&c_{\phi}(\alpha(x)-
\alpha(|\zeta_{\text{ini}})|)\int_{|E_0(\zeta_{\text{ini}})|}^{x}dy \frac{v_0-
y}{\rho(y)^2}  \nonumber \\ &+& \left. c_{\phi}
\int_{|E_0(\zeta_{\text{ini}})|}^{x}dy \frac{\alpha(y)(v_0-y)}{\rho(y)^2}-
c_{\phi}\int_{|E_0(\zeta_{\text{ini}})|}^{x}\frac{dy}{\rho(y)} \right].
\end{eqnarray}
Now we need to compute
\begin{eqnarray}
&&\int_0^{E^+} dy \psi_{\sigma}\left(E_0^{-1}(-
y)\right)=\psi_{\sigma}(\zeta_{\text{ini}})v_0 \nonumber \\ &-&
c_{\phi}\int_0^{E^+} dy\int_{|E_0(\zeta_{\text{ini}})|}^{y}dx \frac{v_0-
x}{x\rho(x)^2} \nonumber \\ &-&c_{\sigma \rho}\int_0^{E^+}
dy\int_{|E_0(\zeta_{\text{ini}})|}^{y}dx \frac{(v_0-x)\alpha(x)}{x\rho(x)}  \\
&+&\frac{c_{\phi}}{v_0}\int_0^{E^+} dy\int_{|E_0(\zeta_{\text{ini}})|}^{y}dx
\frac{(v_0-x)\alpha(x)}{x\rho(x)}\int_0^x dz\frac{v_0-z}{\rho(z)^2} \nonumber
\end{eqnarray}
The repeated integrals can be computed by parts to get
\begin{eqnarray}
\label{eqtemp1}
&&\int_0^{E^+} dx \psi_{\sigma}\left(E_0^{-1}(-
x)\right)=\psi_{\sigma}(\zeta_{\text{ini}})v_0 \nonumber \\ &-& c_{\phi}v_0
\int_{|E_0(\zeta_{\text{ini}})|}^{E^+}dx \frac{v_0-x}{x\rho(x)^2}
+c_{\phi}\int_0^{E^+} dx \frac{v_0-x}{\rho(x)^2} \nonumber \\ &-&c_{\sigma
\rho}v_0 \int_{|E_0(\zeta_{\text{ini}})|}^{E^+}dx \frac{(v_0-
x)\alpha(x)}{x\rho(x)}  \nonumber \\ &+&c_{\sigma \rho}\int_0^{E^+} dx
\frac{(v_0-x)\alpha(x)}{\rho(x)}  \\ &+&c_{\phi}
\int_{|E_0(\zeta_{\text{ini}})|}^{E^+}dx \frac{(v_0-
x)\alpha(x)}{x\rho(x)}\int_0^x dy\frac{v_0-y}{\rho(y)^2} \nonumber \\ &-
&\frac{c_{\phi}}{v_0}\int_0^{E^+} dx \frac{(v_0-x)\alpha(x)}{\rho(x)}\int_0^x
dy\frac{v_0-y}{\rho(y)^2} \nonumber
\end{eqnarray}
For the first integral in Eq.~(\ref{eqtemp1}), we use Eq.~(\ref{psis-zetain}) in
order to extract the part which diverges as $\zeta_{\text{ini}} \to -\infty$.
Now we substitute Eq.~(\ref{eqtemp1}) in Eq.~(\ref{btemp}), and in the final
result, in order to eliminate a maximum of double integral, we use the following
relation
\begin{eqnarray}
    &&\int_0^{E^+} dx \frac{v_0-
x}{\rho(x)}\int_{|E_0(\zeta_{\text{ini}})|}^{x}dy \frac{\alpha(y)(v_0-
y)}{\rho(y)^2} \nonumber \\ &=& \int_0^{E^+} dx \frac{(v_0-x)^2}{\rho(x)^2} -
\frac{v_0-|E_0(\zeta_{\text{ini}})|}{\rho_0(\zeta_{\text{ini}})} \int_0^{E^+} dx
\frac{v_0-x}{\rho(x)}  \nonumber \\ &+&\int_0^{E^+} dx \frac{v_0-
x}{\rho(x)}\int_{|E_0(\zeta_{\text{ini}})|}^x\frac{dy}{\rho(y)}
\end{eqnarray}
The final expression for $B$ is then given by
\begin{eqnarray}
\label{expressionB}
B&=&\frac{c_{\phi}v_0^2}{\sigma^-
}\zeta_{\text{ini}}+v_0^2\psi_{\sigma}(\zeta_{\text{ini}}) -
c_{\zeta_{\text{ini}}}\int_0^{E^+} dx \frac{v_0-x}{\rho(x)} \nonumber \\
&-&c_{\phi}v_0^2\int_{|E_0(\zeta_{\text{ini}})|}^{E^+}dx\left[\frac{1}{\rho(x)}-
\frac{1}{\sigma^-} \right]\frac{v_0-x}{x\rho(x)} \nonumber \\
&+&c_{\phi}v_0\int_{0}^{E^+}dx \frac{v_0-x}{\rho(x)^2}
-c_{\sigma \rho}v_0^2\int_{|E_0(\zeta_{\text{ini}})|}^{E^+}dx \frac{(v_0-
x)\alpha(x)}{x\rho(x)} \nonumber \\
&+&c_{\phi}v_0\int_{|E_0(\zeta_{\text{ini}})|}^{E^+} dx \frac{(v_0-
x)\alpha(x)}{x\rho(x)}\int_0^{x}dy \frac{(v_0-y)}{\rho(y)^2} \nonumber \\
&+&\frac{c_{\phi}(v_0-
|E_0(\zeta_{\text{ini}})|)}{\rho_0(\zeta_{\text{ini}})}\int_{0}^{E^+}dx
\frac{v_0-x}{\rho(x)} \nonumber \\
&-&c_{\phi}\int_{0}^{E^+}dx \frac{(v_0-x)^2}{\rho(x)^2}+c_{\sigma
\rho}v_0\alpha(|\zeta_{\text{ini}}|)\int_0^{E^+}dx\frac{v_0-x}{\rho(x)}
\nonumber \\
&-&c_{\phi}\alpha(|\zeta_{\text{ini}}|)\int_{0}^{E^+} dx \frac{(v_0-
x)}{\rho(x)}\int_{|E_0(\zeta_{\text{ini}})|}^{x}dy \frac{(v_0-y)}{\rho(y)^2}
\nonumber \\
&+&c_{\phi}\int_{0}^{E^+} dx \frac{(v_0-
x)\alpha(x)}{\rho(x)}\int_{|E_0(\zeta_{\text{ini}})|}^{x}dy \frac{(v_0-
y)}{\rho(y)^2} \nonumber \\
&-&c_{\phi}\int_{0}^{E^+} dx \frac{(v_0-x)\alpha(x)}{\rho(x)}\int_{0}^{x}dy
\frac{(v_0-y)}{\rho(y)^2}.
\end{eqnarray}
This expression can be written as in the main text
\begin{equation}
    B=\frac{c_{\phi}v_0^2}{\sigma^-}\zeta_{\text{ini}}+ a_B+b_B e^{\lambda
\zeta_{\text{ini}}},
\end{equation}
where
\begin{eqnarray}
\label{express-aB}
a_B&=&v_0^2\psi_{\sigma}(\zeta_{\text{ini}}) -
\tilde{c}_{\zeta_{\text{ini}}}\int_0^{E^+} dx \frac{v_0-x}{\rho(x)} \nonumber \\
&-&c_{\phi}v_0^2\int_{0}^{E^+}dx\left[\frac{1}{\rho(x)}-\frac{1}{\sigma^-}
\right]\frac{v_0-x}{x\rho(x)} \nonumber \\
&+&c_{\phi}v_0\int_{0}^{E^+}dx \frac{v_0-x}{\rho(x)^2}
-c_{\sigma \rho}v_0^2\int_{0}^{E^+}dx \frac{(v_0-x)\alpha(x)}{x\rho(x)}
\nonumber \\
&+&c_{\phi}v_0\int_{0}^{E^+} dx \frac{(v_0-x)\alpha(x)}{x\rho(x)}\int_0^{x}dy
\frac{(v_0-y)}{\rho(y)^2} \nonumber \\
&+&\frac{c_{\phi}v_0}{\sigma^-}\int_{0}^{E^+}dx \frac{v_0-x}{\rho(x)} -
c_{\phi}\int_{0}^{E^+}dx \frac{(v_0-x)^2}{\rho(x)^2},
\end{eqnarray}
with $\tilde{c}_{\zeta_{\text{ini}}}$ defined by Eq.~(\ref{czetatilde}).

\end{document}